\documentclass{elsart}
\usepackage{amssymb}
\usepackage{amsmath}
\usepackage{mathrsfs}
\usepackage{epsfig,graphicx}
\usepackage{wrapfig}
\usepackage{array}
\usepackage{rotating}

\newcommand{\mrs}{\mathscr}
\newcommand{\vecb}[1]{\mbox{\boldmath$#1$}}
 
\begin{document} 
 
\begin{frontmatter}

\title{Loosely bound three-body nuclear systems in~the $J$-matrix approach}
\author[Lur]{Yu. A. Lurie} and
\ead{ylurie@yosh.ac.il}
\author[AMS]{A. M. Shirokov}\ead{shirokov@nucl-th.sinp.msu.ru}
\address[Lur]{The College of Judea and Samaria, Ariel 44837, Israel}
\address[AMS]{Skobeltsyn Institute of Nuclear Physics, Moscow State
University, Moscow, 119992, Russia}
\begin{abstract}
We  discuss the extension of the oscillator-basis
$J$-matrix formalism on the case of true $A$-body scattering. The
formalism is applied to loosely-bound $^{11}$Li and $^6$He nuclei
within  three-body  cluster models ${\rm {^9Li}}+n+n$ and
$\alpha+n+n$. The  $J$-matrix formalism is used not only
for the calculation of the three-body continuum spectrum wave functions 
but also for the calculation of the
$S$-matrix poles associated with the   $^{11}$Li and $^6$He ground
states to improve the description of the binding energies and ground
state properties. 
The effect of the phase equivalent transformation of the $n{-}\alpha$
interaction  on the properties of the $^6$He
nucleus is examined. 
\end{abstract}
\end{frontmatter}

\section{Introduction}

In this paper we discuss a unified $J$-matrix approach to many-body
systems. The approach links the quantum scattering theory formalism
with traditional variational methods of nuclear theory based
on the wave function expansion  in the harmonic oscillator function
series. The formalism  extends
the variational methods on the case of the continuum spectrum
states. On the other hand, the variational description of the discrete
spectrum states is improved by the use of the 
methods of the scattering theory. The $J$-matrix approach is supposed
to be a valuable extension of the shell model and other nuclear
structure models. Below we examine in detail the application of the
$J$-matrix approach to the light weakly bound nuclei $^{11}$Li and
$^6$He  within the three-body cluster model.

Weakly bound nuclei are extensively studied now both theoretically and
experimentally. New methods are developed in the nuclear structure
theory providing accurate description of the charge and matter distribution
in the nuclear halo, allowing for the coupling to the continuum
spectrum states, etc. At the same time, the traditional language of
nuclear theory, i.~e. the nuclear shell model, is  based on the
oscillator basis expansion of the wave functions and seems to be
inadequate for the the continuum or weakly bound states with the  wave
functions slowly decreasing at large distances since the oscillator
functions are  rapidly decreasing asymptotically. We however
believe that the traditional oscillator-basis nuclear theory can be
successfully developed to face the above challenges by implementing the
$J$-matrix formalism. 

Within the $J$-matrix formalism, the continuum spectrum wave function
 is expanded in infinite series of $L^2$ functions.
 The $J$-matrix method was initially proposed in atomic physics
\cite{Hell,J-matrix-theory-Yamani} and  shown  to be one of
the most efficient and precise methods in calculations of
photoionization \cite{Broad,St-Uzh,St-izv} and electron scattering by atoms
\cite{Konov}. In nuclear physics the same approach has been developed
independently \cite{Fil,NeSm} as the method of harmonic oscillator
representation of scattering theory. This method  has been
successfully used in various  nuclear applications allowing for
 two-body continuum channels, e.~g. 
nucleus-nucleus scattering  has been studied in the algebraic
version of RGM based on  the $J$-matrix formalism  (see the review papers
\cite{FVCh,Rev}); the effect of $\Lambda$ and neutron decay channels in
hypernuclei production reactions has been investigated in Refs.
\cite{M1,M2}, etc.

In this paper we study
exotic neutron-excess nuclei  $^{11}$Li and $^6$He in the
three-body cluster models  
${\rm {^{11}Li}={^9Li}}+n+n$ and ${\rm {^{6}He}}=\alpha+n+n$.
The
two-neutron separation energy in both of these nuclei is small
compared with the neutron separation energy in the $^9$Li and
$^4$He clusters; as a result the wave function of the pair of neutrons
decreases 
slowly with distance and the rms radius of the two-neutron distribution is
large compared with the rms radius of the $^9$Li or $^4$He core (the
so-called {two-neutron halo}).  We note that $^{11}$Li and $^6$He are the
so-called  {Borromean nuclei}, i.~e. none of the two-body
subsystems in the cluster decompositions ${\rm {^{11}Li}={^9Li}}+n+n$
and ${\rm {^{6}He}}=\alpha+n+n$ has a bound state. We employ the
three-body extension of the $J$-matrix formalism for the description
of the three-body decay channels and develop this approach to improve
the description of the  ground state properties.

Three-body decay channels are much more informative than the two-body
ones, however the complete  description of a three-body decay is a
very complicated problem. Nevertheless in some 
cases, one can use the so-called
{democratic decay approximation} for the successful analysis of
the experimental data or for obtaining reliable theoretical predictions.

Generally, a three-body  continuum spectrum wave function in the
asymptotic region 
is a superposition of the components describing two-
and three-body decay channels~\cite{MeFa}.  The democratic decay approximation
accounts for the three-body channel only;  two-body channels associated with
the appearance of bound two-body subsystems and `{\em non-democratic}'
subdivision of the system of the type (2+1), are not allowed for in the
approximation. Therefore, this approximation is valid for the study of a
three-body system  only in the case when all two-body decay channels are closed
and the only open channel is a three-body one; in other words, this is
the case when none of the two-body subsystems has a bound state. 
Hence the democratic decay approximation is adequate for  the study of
Borromean nuclei $^6$He and $^{11}$Li at least at small enough excitation
energies (less than the single nucleon binding energy
in the $^9$Li and $^4$He clusters).

Generally speaking, if
the three-body channel is the only open one, than the system can have
two types of asymptotics~\cite{MeFa}. One of the asymptotics corresponds to the
situation when one of the particles is scattered by another  and the
third particle is a spectator.
From the general physical point of view,
this asymptotics is supposed to be of little importance for
a nuclear system excited in some nuclear reaction and decaying via a
three-body channel. The alternative type of the asymptotics
is a superposition of ingoing and outgoing six-dimensional spherical waves,
it corresponds to the situation when the decaying system emits
(or/and absorbs) 
three particles from  some point in space. Allowing for the three-body
asymptotics of this type  only is a quintessence of the democratic
decay approximation; sometimes this approximation is also referred to
as {\em true three-body scattering} or  
{\em $\mbox{3} \to\mbox{3}$ scattering}.

Nuclear structure studies in the framework of the democratic decay
approximation have been started in 1970th by R.~I.~Jibuti and
collaborators.  The review of the results obtained by the Tbilisi group can
be found in Refs.~\cite{Jib-EChAYa,JiKr}. Recently a
considerable progress has been made in both 
theoretical and experimental studies of democratic decays (see reviews
in  Refs.~\cite{DaZhu-rep,Hanoi,DaZhu-rep2,Fed-rep}). 

The generalization of the $J$-matrix formalism on the case of $3\to 3$
scattering (and on a more general  case of $N\to N$ scattering) was
suggested in Ref.~\cite{SmSh}.  First it was successfully applied in
the study of
monopole excitations in $^{12}$C nucleus in the cluster model 
${\rm {^{12}C}}=\alpha+\alpha+\alpha$ in Ref.~\cite{Mikh}. Later by
means of this approach we studied  $\Lambda\Lambda$ hypernuclei
\cite{Zeit}.  Our first attempts to employ this approach in the
studies of
$^{11}$Li and $^6$He nuclei  in the three-body cluster models  
can be found in Refs.~\cite{Izv1,Halo,Mex,PHT}.

We use the $J$-matrix $3\to 3$ scattering  formalism for the
construction of the three-body continuum spectrum wave functions of 
 $^{11}$Li and $^6$He nuclei excited in nuclear reactions. However, probably 
more interesting is the application of this formalism to the
study of the ground states of these nuclei treated as three-body
systems ${\rm {^9Li}}+n+n$ and $\alpha+n+n$.  The idea is the
following. By the $J$-matrix formalism we calculate the $3\to 3$
scattering $S$-matrix. This calculation may be extended on the complex
momentum plane. Hence we can locate numerically the $S$-matrix poles
associated with the bound states.  The $S$-matrix pole calculations
improve the variational results for binding energies obtained by the pure
diagonalization of the truncated Hamiltonian matrix. Knowing the
binding energies, we calculate the bound state wave functions by means
of the $J$-matrix formalism. We suppose that the approach based on the
$S$-matrix pole calculation will be useful not only within
the cluster models; the implementation of the $A\to A$ scattering $S$-matrix
calculation within the $J$-matrix formalism is believed to improve
essentially the traditional nuclear shell model.

The first calculation of the $S$-matrix poles within the $J$-matrix
formalism was performed in Ref.~\cite{OkhrS} where resonances in
$\alpha{-}\alpha$ scattering were investigated. The  $S$-matrix poles
in He atom and H$^-$ ion were calculated in Ref.~\cite{St-Uzh,St-izv}
in the $J$-matrix model suggested by Broad and
Reinhardt~\cite{Broad}. The resonance energies and widths 
were reproduced with a high
accuracy. The  $S$-matrix poles associated with the bound states were
also calculated in  Ref.~\cite{St-Uzh,St-izv}. The binding energies
were reproduced with a very high accuracy; we note also that a very
large number of bound states was obtained by means of the $S$-matrix
pole  calculations: it was not only much larger than the number of
bound states obtained variationally (by the diagonalization of the
truncated Hamiltonian matrix), but even much larger than the
rank of the truncated Hamiltonian matrix. 
This approach not only provides
an essential improvement in calculations of binding energies, but also
makes it possible to 
calculate the ground state wave function with the correct
asymptotics. It is  very important for halo nuclei like $^{11}$Li
and $^6$He due to the slow decrease of the wave function in the
asymptotic region that cannot be reproduced by a superposition of a
finite number of oscillator functions obtained in variational
calculation. As a result, the rms radius, electromagnetic transition
probabilities and other observables are improved essentially.

We also discuss a phase-equivalent transformation
suggested 
in Ref.~\cite{PHT} that  
is used to obtain families of
phase-equivalent  $n{-}\alpha$ potentials. The effect of the
phase-equivalent transformation of the $n{-}\alpha$ interaction on the
$^6$He properties is examined. 

We start the discussion with a brief sketch of the $N\to N$ scattering
$J$-matrix formalism  suggested in Ref.~\cite{SmSh}.

\section{$J$-matrix formalism with hyperspherical oscillator basis}

The wave function $\Psi$ of a system of $A$ particles is generally a
function of $A$ 
coordinates of individual particles $\mbox{\boldmath$r$}_i$. The
center of mass motion can be separated and eliminated; 
as a result we can treat the wave function $\Psi$ as a
function of $A-1$
Jacobi coordinates $\mbox{\boldmath$\xi$}_i$. Within the democratic
decay approximation, it is natural to employ the hyperspherical
harmonics formalism (see, e.~g., \cite{JiKr,29}).  The
$3A-3$ independent variables  $\mbox{\boldmath$\xi$}_i$ are
rearranged in this formalism:  the so-called hyperradius
$\rho = \sqrt{\sum\limits_{i=1}^{A-1} \mbox{\boldmath$\xi$}_i^2} $
is introduced and all the rest $3A-4$ variables are the angles on the
$(3A-3)$-dimensional sphere. The wave function is searched for in the form
\begin{gather}
\Psi = \sum_{\Gamma} d_{\Gamma}
\Psi_{\Gamma},
\label{e21}
\intertext{where}
\Psi_\Gamma\equiv\Psi_{K\gamma}= 
                    \Phi_{K\gamma}(\rho)\;{\mrs Y}_{K \gamma}(\Omega),
\label{hyper-K}
\end{gather}
$\Omega$ is the set of angles on the
$(3A-3)$-dimensional sphere,  $K$ is the
so-called hypermomentum [physically $K$ is the angular momentum in the
$(3A-3)$-dimensional space; $K\geq L$ where $L$ is the orbital angular
momentum] and $\gamma$ stands for all the rest
quantum numbers labelling the   hyperspherical function  
${\mrs Y}_{K \gamma}(\Omega)$, the multi-index $\Gamma=\{K,\gamma\}$. 
Various analytic expressions may be
found for the   hyperspherical functions  
${\mrs Y}_{K \gamma}(\Omega)$ in textbooks (see, e.~g., \cite{JiKr}).

In the two-body case, 
$\rho=r=|\mbox{\boldmath$r$}_1-\mbox{\boldmath$r$}_2|$ is the distance
between the particles and ${\mrs Y}_{K \gamma}(\Omega)$ becomes the
usual spherical function $Y_{LM}(\Omega)$ with $K$ and $\gamma$
playing the role of the angular momentum $L$ and its projection $M$,
respectively. Therefore equation (\ref{hyper-K}) is a generalization
on the $A$-body  ($A\geq 3$) case of the conventional 
wave function used in the study of spherically symmetric two-body
systems. However, contrary to the case of two-body systems with
central interaction which impose the spherical symmetry, the
hypermomentum $K$ is not an integral of motion:    the two-body
interactions in the $A$-body system couples the states with different
values of the hypermomentum  $K$ and we  have a set of coupled
equations for the hyperradial functions $\Phi_{K\gamma}(\rho)$. This set of
equations  has the same structure as the one 
describing 
two-body systems with non-central forces; from the point of view of
scattering theory, this set of coupled equations is formally
equivalent to the one describing
multichannel scattering in the system with the hyperspherical channels
$\Gamma$.

We introduce the hyperspherical oscillator basis
\begin{equation}
|\kappa\Gamma\rangle\equiv |\kappa K\gamma\rangle=
      \mrs R_{\kappa K}(\rho)\: {\mrs Y}_{K \gamma}(\Omega),
\label{HHO-basis}
\end{equation}
where the hyperradial oscillator function
\begin{align}
&\mrs R_{\kappa K}(\rho) \equiv \mrs R_\kappa^{\mrs L}(\rho)=
  \rho^{-(3A-4)/{2}}\: \frak r
_{\kappa K}(\rho),
\label{e23} \\[2ex]
&\frak r
_{\kappa K}(\rho) \equiv \frak r
_\kappa ^{\mrs L}(\rho)=(-1)^\kappa\
\sqrt{ \frac{2 
\kappa!}{\Gamma(\kappa + {\mrs L} +  3/2) } }\ \
               \rho^{ {\mrs L} +1}\:
             {e}^{-
                       \rho^2/2}\:
        L_\kappa^{{\mrs L}+\frac12}(
                                         \rho^2),
\label{e24}
\end{align}
{$L^{\alpha}_n(x)$ is the associated Laguerre polynomial and}
\begin{equation}
{\mrs L} = K + \frac{\displaystyle 3A-6}{\displaystyle 2}.
\label{calL}
\end{equation}
In the two-body case ($A=2$), ${\mrs L}$ is equivalent to the orbital angular
momentum $L$, ${\mrs L}=L$, and Eqs.~(\ref{HHO-basis})--(\ref{calL})
define the conventional oscillator basis. In the many-body case
($A\geq 3$), Eqs.~(\ref{HHO-basis})--(\ref{calL})
define the $A$-body oscillator functions, i.~e. the eigenfunctions of
the $A$-body Schr\"odinger equation with the potential energy
\begin{equation} 
U = \frac{\omega^{2}}{2}\: \sum_{i=1}^A\: m_i\, (\vecb{r}_{i}-\vecb{R})^{2} 
                = \frac{\hbar \omega}{2} \,  \rho^{2} ,
\end{equation} 
where $m_i$ is the mass of the $i$th particle, $\vecb{R}$ is the
center-of-mass coordinate and $\omega$ is the parameter of the
oscillator basis. The corresponding   eigenenergy is 
\begin{equation}
E_{\kappa K}=\left(N+\frac{3A-3}{2}\right)\hbar\omega,
\label{eigenHO}
\end{equation}
where the 
number of oscillator quanta
\begin{equation}
N=2\kappa+K=2\kappa +{\mrs L}-\frac{3A-6}{2}.
\label{totalN}
\end{equation}
The hyperspherical oscillator basis (\ref{HHO-basis}) is orthonormalized:
\begin{gather}
\langle \kappa K \gamma | \kappa' K' \gamma' \rangle = \delta_{\kappa\kappa'}
                                                 \delta_{  \Gamma \Gamma'},
\label{e25} 
\intertext{or}
\int\limits_0^\infty \frak r
_{\kappa K}^*(\rho)\frak r
_{\kappa'K}(\rho)\,\mbox{d}\rho 
=\delta_{\kappa\kappa'}.
\label{e25-hr}
\end{gather}

The wave function $\Psi$ is expanded in the hyperspherical oscillator
function series,
\begin{equation}
\Psi=\sum_{\kappa,K,\gamma}
 \langle \kappa K\gamma|\Psi\rangle\; |\kappa K\gamma\rangle ,
\label{hyperr-exp}
\end{equation}
and the Schr\"odinger equation takes the form
\begin{equation}
\sum_{\kappa',K',\gamma'}
\langle\kappa K \gamma | H-E | \kappa' K' \gamma' \rangle\;
\langle \kappa' K' \gamma' | \Psi\rangle = 0,
\label{e26}
\end{equation}
where $E$ is the energy, the Hamiltonian $H=T+V$, the potential energy $V$ is usually a
superposition of two-body interactions $V_{ij}$, 
$V=\sum\limits_{i<j}V_{ij}$, and 
$T$ is the $A$-body kinetic energy operator.
Generally, the Hamiltonian matrix 
$\langle\kappa K \gamma|H| \kappa' K' \gamma' \rangle$ is infinite.
However, within the $J$-matrix formalism we truncate the potential
energy matrix; it is most natural to define the truncation boundary
through the 
number of oscillator quanta, i.~e. we suppose that
\begin{equation}
\langle\kappa K \gamma|V| \kappa' K' \gamma' \rangle=0
\qquad \text{if}\quad 2\kappa+K>\tilde N\quad
\text{or}\quad 2\kappa'+K'>\tilde N.
\label{trunc}
\end{equation}
The kinetic energy matrix $\langle\kappa K\gamma|T|\kappa'K'\gamma'\rangle$ 
is tridiagonal,
\begin{multline}
\langle\kappa K \gamma| T|\kappa' K' \gamma' \rangle =
\frac{\hbar \omega}{2}\;
\delta_{K K'}\:\delta_{\gamma\gamma'}
\left[\:-\ \sqrt{(n+1)\left(n+{\mrs L}+ \frac32
\right)}\ \delta_{\kappa+1,\kappa'}\right.   \\
 + \left(2n + {\mrs L}+ \frac32
\right) \delta_{\kappa\kappa'}  \left. - \
\sqrt{ n\left(n+{\mrs L}+
\frac12\right)}\ \delta_{\kappa-1,\kappa'}
        \right] .
\label{e27}
\end{multline}

The $A$-body
hyperspherical $J$-matrix formalism is very close to the conventional 
$J$-matrix formalism with oscillator basis
\cite{J-matrix-theory-Yamani,Ann}   used in multichannel
two-body problems.  The main difference is that $\mrs L$ entering
Eq.~(\ref{e27}) is half-integer if the number of particles $A$ is odd
[see~(\ref{calL}) and note that $K$ is always integer], in particular,
$\mrs L$ is half-integer in the three-body case discussed here.
The general $J$-matrix oscillator-basis solutions of the free
Schr\"odinger equation
\begin{equation}
\sum_{\kappa',K',\gamma'}
\langle\kappa K \gamma | T-E | \kappa' K' \gamma' \rangle\;
\langle \kappa' K' \gamma' | \Psi\rangle = 0
\label{genJfree}
\end{equation}
that may be used in the case of arbitrary  $\mrs L$, were suggested in
Ref.~\cite{SmSh}. The regular solution is 
\begin{align}
S_{\kappa K}(q)&=
\sqrt{ \frac{2 \kappa!}{\Gamma(n + {\mrs L} + 3/2)} }\;\,
            q^{ {\mrs L} +1}\;
                     {e}^{-q^2/2}\: L_\kappa^{ {\mrs L} + \frac12}(q^2),
\label{e34}    
\intertext{the irregular solutions are}
C_{\kappa K}(q)&=-\, \frac{2q}{\pi S_{0K}(q)}\ 
    \text{\bf V. P.}
\int\limits_0^{\infty}\frac{S_{0K}(q')\:S_{\kappa K}(q')}
{q^2 - {q'}^2}\ {d}q',
\label{e35}\\[2ex]
C_{\kappa K}^{(+)}(q)&=-\, \frac{2q}{\pi S_{0K}(q)}
 \int\limits_0^{\infty}\frac{S_{0K}(q')\:S_{\kappa K}(q')}
{q^2 - {q'}^2 + i0 }\ {d}q',
\label{e36}\\[2ex]
C_{\kappa K}^{(-)}(q)&=-\, \frac{2q}{\pi S_{0K}(q)}
 \int\limits_0^{\infty}\frac{S_{0K}(q')\:S_{\kappa K}(q')}
{q^2 - {q'}^2 - {i}0 }\ {d}q',
\label{e37}
\end{align}
where $\displaystyle q=\sqrt{\frac{2E}{\hbar\omega}}$, {\bf V. P.} in
Eq.~(\ref{e35}) indicates the  principal value integral, $+i0$ and $-i0$ in
Eqs.~(\ref{e36})--(\ref{e37}) show how the poles of the integrand
should be treated. The solutions (\ref{e34})--(\ref{e37}) are not
independent, the following relation between  them is valid:
\begin{equation}
C_{\kappa K}^{(\pm )}(q)\ =\ C_{\kappa K}(q)\ \pm\
          {i}\,S_{\kappa K}(q).
\label{CpmCS}
\end{equation}
However, any pair of the solutions (\ref{e34})--(\ref{e37}) can be
used to construct an arbitrary solution.

The regular solution (\ref{e34}) is just the hyperradial momentum-space
oscillator function.
The irregular solutions  (\ref{e35})--(\ref{e37}) are more
complicated. They were analyzed in detail in Ref.~\cite{SmSh}. In the
general case, they can be expressed through Tricomi function. However
in the case of even $A$, the irregular solutions can be simplified and
expressed through the confluent hypergeometric function. 
The physical
meaning of the solutions   (\ref{e34})--(\ref{e37}) is clear from the
following expressions 
\cite{SmSh}:
\begin{align}
&\sum_{\kappa=0}^{\infty}S_{\kappa K}(q)\, \frak r
_{\kappa K}(\rho)=
\sqrt{q \rho}\:J_{{\mrs L}+\frac{1}{2}}(q \rho)
\mathop{\longrightarrow}\limits_{\rho \to \infty}
\sqrt{ \frac{2}{\pi} }\,
\sin\!\left(q\rho- \frac{\pi {\mrs L}}{2}\right),
\label{e319}\\
&\sum\limits_{\kappa=0}^{\infty} 
          C_{\kappa K}(q)\,\frak r
_{\kappa K}(\rho)
\mathop{\longrightarrow}\limits_{\rho \to \infty}
-\,\sqrt{q\rho}\:N_{{\mrs L}+\frac{1}{2}}(q\rho)
\mathop{\longrightarrow}\limits_{\rho \to \infty} \sqrt{ \frac{2}{\pi} }\:
\cos\!\left(q\rho- \frac{\pi {\mrs L}}{2}\right),
\label{e331} \\
&\sum_{\kappa=0}^{\infty}C_{\kappa K}^{(+)}(q)\,\frak r
_{\kappa K}(\rho)
\mathop{\longrightarrow}\limits_{\rho \to \infty}
\,i\,\sqrt{q \rho}\:H^{(1)}_{{\mrs L}+\frac{1}{2}}(q \rho)
\mathop{\longrightarrow}\limits_{\rho \to \infty}
\sqrt{ \frac{2}{\pi} }\
{e}^{{i} \left(q\rho-\frac{\pi{\mrs L}}{2}\right)},
\label{e330}\\
&\sum_{\kappa=0}^{\infty}C_{\kappa K}^{(-)}(q)\,\frak r
_{\kappa K}(\rho)
\mathop{\longrightarrow}\limits_{\rho \to \infty}
-\,i\,\sqrt{q \rho}\:H^{(2)}_{{\mrs L}+\frac{1}{2}}(q \rho)
\mathop{\longrightarrow}\limits_{\rho \to \infty}
\sqrt{ \frac{2}{\pi} }\
{e}^{-{i} \left(q\rho-\frac{\pi{\mrs L}}{2}\right)}.
\label{e3302}
\end{align}
Here $J_\alpha(x)$, $N_\alpha(x)$ and $H^{(1,2)}_\alpha(x)$ are 
Bessel, Neumann and Hankel functions, respectively.

In the case of continuum spectrum ($E>0$), the oscillator
representation wave function 
$\langle\kappa K\gamma|\Psi^{(\Gamma')}\rangle
    \equiv\langle\kappa\Gamma|\Psi^{(\Gamma')}\rangle$ in 
the channel $\Gamma$  is of the  form
\begin{equation}
\langle \kappa {  \Gamma} | \Psi^{(\Gamma')} \rangle=
\frac12\left(
\delta_{{  \Gamma \Gamma'}}\,C_{\kappa K}^{(-)}(q)  -
C_{\kappa K}^{(+)}(q)\,[{\bf S}]_{{  \Gamma \Gamma'}}
\right)
\label{e49}
\end{equation}
in the `{\em external region}' 
$\kappa\geq\kappa_{\Gamma\vphantom{'}}$, where 
$\kappa_{\Gamma\vphantom{'}}=(\tilde N-K)/2$ is the potential energy
truncation boundary in the channel $\Gamma$.
It is supposed that the ingoing spherical wave is present in the 
channel $\Gamma'$ only while the outgoing spherical waves are 
present in all channels; $[{\bf S}]_{\Gamma\Gamma'}$ entering 
Eq.~(\ref{e49}) is the 
matrix element of the hyperspherical $S$-matrix. The $S$-matrix
$\left[{\bf S}\right]$ 
can be calculated by the following formula \cite{SmSh}:
\begin{equation}
\left[{\bf S}\right] = \left[{\bf A}^s\right]^{-1}\,\left[ {\bf B}^s\right],
\label{e413}
\end{equation}
where the matrix elements of the matrices $\left[{\bf A}^s\right]$
and $\left[{\bf B}^s\right]$ are
\begin{gather}
\left[{\bf A}^s \right]_{{  \Gamma' \Gamma}} =
\langle \kappa_{\Gamma'\vphantom{'}}\Gamma'|{\mrs P}|
\kappa_{\Gamma\vphantom{'}}+1,\Gamma \rangle\,
C_{\!\kappa_{\Gamma\vphantom{'}}+1,\,K}^{(+)}(q)
- \delta_{\Gamma \Gamma'}\,
C_{\!\kappa_{\Gamma\vphantom{'}}K}^{(+)}(q),
\label{e411}\\
\left[{\bf B}^s \right]_{{  \Gamma' \Gamma}} =
\langle \kappa_{\Gamma'\vphantom{'}}\Gamma'|\mrs P|
\kappa_{\Gamma\vphantom{'}}+1,\Gamma\rangle\,
C_{\!\kappa_{\Gamma\vphantom{'}}+1,\,K}^{(-)}(q)
- \delta_{\Gamma\Gamma'}\,
C_{\!\kappa_{\Gamma\vphantom{'}}K}^{(-)}(q);
\label{e412}
\end{gather}
the matrix elements of the matrix $\left[\mrs P\right]$  
proportional to the discrete
analog of the $P$-matrix (see Ref.~\cite{Ann} for details) are 
\begin{equation}
\langle \kappa\Gamma |\mrs P|\kappa_{\Gamma'\vphantom{'}}+1,\Gamma'\rangle =
\langle\kappa\Gamma |{\frak P}|\kappa_{\Gamma'\vphantom{'}} \Gamma' \rangle\,
\langle\kappa_{\Gamma'\vphantom{'}}\Gamma'|T|
\kappa_{\Gamma'\vphantom{'}}+1,\Gamma' \rangle,
\label{e43}
\end{equation}
the kinetic energy matrix elements 
$\langle\kappa_{\Gamma'\vphantom{'}}\Gamma'| T|
   \kappa_{\Gamma'\vphantom{'}}+1,\Gamma' \rangle$
are given by Eq.~(\ref{e27}), and the matrix  
$\left[{\frak P}\right]\equiv \left[H-E\right]^{-1}$
defined in the truncated model space spanned by oscillator functions
(\ref{HHO-basis}) with 
$\kappa\leq \kappa_{\Gamma\vphantom{'}}=(\tilde N-K)/2$ in each
channel $\Gamma$, has the matrix elements 
\begin{equation}
\langle\kappa\Gamma|{\frak P}|\kappa'\Gamma' \rangle =
\sum_{\lambda} \frac{\langle\kappa\Gamma|\lambda \rangle
\langle\lambda|\kappa'\Gamma' \rangle }
{E - E_{\lambda} } .
\label{e44}
\end{equation}
In Eq.~(\ref{e44}), $E_\lambda$ are eigenenergies and
$\langle\kappa\Gamma|\lambda \rangle$ are the respective eigenvectors of
the truncated Hamiltonian matrix, i.~e.  $E_\lambda$  and 
$\langle\kappa\Gamma|\lambda \rangle$ can be found by solving Eq.~(\ref{e26})
supposing that $\kappa\leq\kappa_{\Gamma\vphantom{'}}$ and 
$\kappa'\leq\kappa_{\Gamma{'}\vphantom{'}}$. The results obtained by
diagonalization of the truncated Hamiltonian matrix we shall refer to
as {\em variational results}.

After performing variational calculation, we obtain the $S$-matrix
with the help of Eqs.~(\ref{e413})--(\ref{e44}) and  the oscillator
representation wave function  $\langle\kappa\Gamma|\Psi^{(\Gamma')}\rangle$
in the external region 
$\kappa\geq\kappa_{\Gamma\vphantom{'}}$ by the formula
(\ref{e49}). The  oscillator representation wave function
$\langle\kappa\Gamma|\Psi^{(\Gamma')}\rangle$ in the `{\em internal region}'
$\kappa\leq\kappa_{\Gamma\vphantom{'}}$ can be now calculated as
\begin{equation}
\langle\kappa\Gamma|\Psi^{(\Gamma')}\rangle=
\sum_{\Gamma''}\; \langle\kappa \Gamma|\mrs P|
        \kappa_{\Gamma''\vphantom{'}}+1,\Gamma'' \rangle\,
\langle\kappa_{\Gamma''\vphantom{'}}+1,\Gamma''| \Psi^{(\Gamma')}\rangle,
   \qquad \kappa\leq \kappa_{\Gamma\vphantom{'}}.
\label{e42}
\end{equation}

Equation (\ref{e413}) can be used in the complex momentum plane; the
bound state energies are associated with the $S$-matrix poles that can
be found by the numerical solution of the obvious equation
\begin{equation}
\det \left[{\bf A}^s\right] =0.
\label{Poles}
\end{equation}
The matrix $\left[{\bf A}^s\right]$ in Eq.~(\ref{Poles}) is the
extension on the 
complex momentum plane of~(\ref{e411}); for the the bound states with
$E=\frac12 q^2\hbar\omega<0$ its matrix elements are
\begin{equation}
\left[{\bf A}^s \right]_{{  \Gamma' \Gamma}} =
\langle \kappa_{\Gamma'\vphantom{'}}\Gamma'|{\mrs P}|
\kappa_{\Gamma\vphantom{'}}+1,\Gamma \rangle\,
C_{\!\kappa_{\Gamma\vphantom{'}}+1,\,K}^{(b)}(q)
- \delta_{\Gamma \Gamma'}\,
C_{\!\kappa_{\Gamma\vphantom{'}}K}^{(b)}(q),
\label{boundA}
\end{equation}
where 
\begin{equation}
C_{\kappa K}^{(b)}(q)=
C_{\kappa K}^{(+)}(i|q|) .
\label{C-bound}
\end{equation}
The relation (\ref{e49}) for the oscillator representation wave function in 
the external region $\kappa\geq\kappa_{\Gamma\vphantom{'}}$ should be
replaced for bound states by
\begin{equation}
\langle \kappa {  \Gamma} | \Psi \rangle=  
-[{\bf S}]_{\Gamma\Gamma}\, C_{\kappa K}^{(b)}(q);
\label{boundOWF}
\end{equation}
the multipliers $[{\bf S}]_{\Gamma\Gamma}$ are obtained by the
numerical solution of the equation
\begin{equation}
[{\bf S}]_{{\Gamma \Gamma}}\,
 C_{\!\kappa_{\Gamma\vphantom{\Gamma'}}K}^{(b)}(q)=
\sum_{{\Gamma'}}\; \langle\kappa_{\Gamma\vphantom{'}}\Gamma|\mrs P|
\kappa_{\Gamma'\vphantom{'}}+1,\Gamma' \rangle\;
[{\bf S}]_{{  \Gamma' \Gamma'}}\,
C_{\!\kappa_{\Gamma'\vphantom{'}}+1,\,K}^{(b)}(q)
\label{e46}
\end{equation}
[we note that Eq.~(\ref{Poles}) is the condition of solvability of
Eq.~(\ref{e46}) for the bound states]. Equation~(\ref{e42}) can be
used for the calculation of the bound state  oscillator representation wave
function $\langle\kappa\Gamma|\Psi^{(\Gamma')}\rangle$ in 
the internal region $\kappa\leq\kappa_{\Gamma\vphantom{'}}$. Since the
set of  $[{\bf S}]_{\Gamma\Gamma}$ can be obtained from
Eq.~(\ref{e46}) up to a common multiplier only, the bound state wave
function should be normalized numerically.

The bound state  energies and wave functions obtained by numerical
calculation of the $S$-matrix poles, will be refered to as {\em
$J$-matrix results}.

In practical applications,  it is useless to allow for all possible
channels $\Gamma=\{K,\gamma\}$ in the external region 
$N\equiv 2\kappa+K>\tilde N$. We start the calculations
allowing for the channels 
$\Gamma$ with few minimal possible values of $K$ in the external
region only, i.~e. we allow for the  channels with  all possible values of
$K\leq K_{\rm tr}$ and all possible $\gamma$ values for a given $K$ in
the external region.
Therefore the summation over $\Gamma''$ in (\ref{e42}) is
restricted to these allowed channels. However all possible channels
with $K\leq \tilde N$ are allowed for in the internal region 
$N\equiv 2\kappa+K\leq\tilde N$. As a next step, we increase the value
of $ K_{\rm tr}$ and allow for more
channels $\Gamma$ in the external region. The convergence of all
results (binding energies and other bound state observables,
transition probabilities, etc.)  with $ K_{\rm tr}$ is carefully
examined. The convergence is usually achieved at small enough values
of $ K_{\rm tr}$ (much less than $\tilde N$). Such converged with
respect to $ K_{\rm tr}$ results are discussed below.

\section{Application to $\bf^{11}Li$ nucleus} 
 The $^{11}$Li nucleus is studied in the three-body cluster model 
${\rm {^9Li}}+n+n$.
The three-body cluster model for $^{11}$Li
 was first suggested in Ref.~\cite{Hansen-Jensen}. The $^{11}$Li
 three-body binding energy or equivalently the $^{11}$Li two-neutron
 separation energy  is 
$0.295\pm 0.035$~MeV~\cite{Li11-energy-experiment-Benenson} that is
 much less than the excitation energy of the lowest excited state in
 the cluster $^9$Li. As it was already noted,  $^{11}$Li is a Borromean
 nuclei, i.~e. none of the two-body subsystems ${\rm{^9Li}}+n$ and
 $n+n$ has a bound state.

The two-body potentials are needed for the investigation of the
three-body system ${\rm{^{11}Li}}={{\rm{^9Li}}+n+n}$. Unfortunately
the information about the $n{-}{\rm{^9Li}}$ interaction is scarce. As
a result, in Ref.~\cite{Hansen-Jensen} the $n{-}{\rm{^9Li}}$ potential
was phenomenologically parametrized and fitted to the  $^{11}$Li
binding energy. Therefore it is reasonable to use a 
simplified
three-body model with simplified interactions to avoid computational
difficulties; a microscopic extension of the model seems to be useless
due to uncertainties of the $n{-}{\rm{^9Li}}$ interaction.

We employ the $n{-}n$ and $n{-}{\rm{^9Li}}$ potentials of
Ref.~\cite{Hansen-Jensen}. The Gaussian $n{-}n$ potential
\begin{equation}
U(r) = -V_0 \,  e^{- \left( r / R \right)^2 }
\label{GaussPotential}
\end{equation}
was fitted to the low-energy singlet ($S=0$) $s$ wave phase shifts, its depth 
 $V_0 =31$~MeV and its width $R=1.8$~fm. The $n{-}{\rm{^9Li}}$
 potential is of a two-Gaussian form,
\begin{equation}
U(r) = -V_1 \,  e^{- \left( r / R_{1} \right)^2 }
     -V_2 \,  e^{- \left( r / R_{2} \right)^2 } ,
\label{twoGaussPotential}
\end{equation}
with $V_1=7$~MeV, $R_1=2.4$~fm,  $V_2=1$~MeV, and $R_2=3.0$~fm.
It is supposed that two valent neutrons in $^{11}$Li are
in a singlet spin state. The total angular momentum $J$ in
$^{11}$Li  results from the coupling of the  spin $\frac32$ of the
$^9$Li cluster  with the relative motion orbital angular
momentum $L$; however we do not make use of the spin-flip operators in
our investigation, therefore we can disregard the spin variables and
exclude them from the wave function.
The  $^{11}$Li ground state orbital angular
momentum $L$ is supposed to be zero. We can define the channels
$\Gamma$ as $\Gamma = \{ K, l_x, l_y, L \}$ where $l_x$ and  $l_y$ are
orbital angular momenta corresponding to the  Jacobi coordinates
\begin{align}
 \vecb{x} &=
 \sqrt{ \frac{\omega}{\hbar} \, 
                        \frac{m_{1}m_{2}}{m_{1}+m_{2}} } 
                \ (\vecb{r}_{1} - \vecb{r}_{2}) , \label{Jac-x} \\ 
 \vecb{y} &=
 \sqrt{ \frac{\omega}{\hbar} \, 
                    \frac{m_{3}(m_{1}+m_{2})}{m_{1}+m_{2}+m_{3}} } 
     \ \left( 
     \frac{ m_{1}\vecb{r}_{1} + m_{2}\vecb{r}_{2} } { m_{1} + m_{2} } 
                               - \vecb{r}_{3}  \right)\!; \label{Jac-y}
\end{align}
the angular momenta $l_x$ and  $l_y$ are coupled to the orbital angular
momentum $L$.
The wave function is given by the general formula
(\ref{hyperr-exp}). The oscillator basis (\ref{HHO-basis}) in our case
takes the form
\begin{gather}
|\kappa\Gamma\rangle\equiv | \kappa K l_{x} l_{y} : L M\rangle
\equiv\langle \mbox{\boldmath$\rho$} | \kappa K l_{x} l_{y} : L M\rangle = 
  \mrs R_{\kappa, K}(\rho) 
  \  \mrs Y_{K \gamma}(\Omega), 
                   \label{11Li-basis} 
\end{gather}
where the hyperspherical functions
\begin{multline}
{\mrs Y}_{K \gamma}(\Omega)\equiv 
\mrs Y_{K l_{x} l_{y} L M}(\Omega)  
\\=
N^{l_{x} l_{y}}_{K} \, \cos^{l_{x}}\!\alpha \, \sin^{l_{y}}\!\alpha
           \ {P}^{l_{y}+1/2, \, l_{x}+1/2}_{n}(\cos 2\alpha)
\\ \times 
\sum_{m_{x}+m_{y}=M} 
  (l_{x} m_{x} l_{y} m_{y} | L M ) 
  \; Y_{l_{x}, m_{x}}(\hat{\vecb{x}}) \; Y_{l_{y}, m_{y}}(\hat{\vecb{y}}), 
                   \label{hyp_ang_fun} 
\end{multline}
$\alpha = \tan^{-1}(y/x)$,
$Y_{lm}$ is the spherical function, $(l_{x} m_{x} l_{y} m_{y}|LM)$ is 
the Clebsch--Gordan coefficient, $P_n^{\alpha,\beta}(x)$ is Jacobi
polynomial \cite{Abr}, and the normalization factor 
$$
N^{l_{x} l_{y}}_{K} = \sqrt{ 
\frac{2 n! \, (K+2) \, (n+l_{x}+l_{y})!} 
          {\Gamma(n+l_{x}+3/2) \, \Gamma(n+l_{y}+3/2)} } \:.
$$

The basis (\ref{11Li-basis}) is useful for the calculation of
three-body decays. However it is not convenient in the calculation of
the matrix elements of two-body potentials $V_{ij}$. These matrix
elements can be easily calculated in the basis
\begin{multline}  
| n_{x} l_{x} n_{y} l_{y} : L M \rangle\equiv
\langle\vecb{x} \vecb{y} | n_{x} l_{x} n_{y} l_{y} : L M \rangle 
\\= 
\sum_{m_{x}+m_{y}=M} 
     (l_{x} m_{x} l_{y} m_{y} | L M ) 
        \, \langle \vecb{x} | n_{x} l_{x} m_{x}\rangle  
          \, \langle \vecb{y} | n_{y} l_{y} m_{y}\rangle, 
                  \label{Jac_fun} 
\end{multline}
where $\langle \vecb{x} | n_{x} l_{x} m_{x}\rangle$ and 
$\langle \vecb{y} | n_{y} l_{y} m_{y}\rangle$ are the conventional
three-dimensional oscillator functions depending on the Jacobi coordinates 
$\vecb{x}$ and $\vecb{y}$, respectively. The unitary transformations
relating the basises (\ref{11Li-basis}) and (\ref{Jac_fun}),
\begin{gather} 
| \kappa K l_{x} l_{y} : L M \rangle = 
   \sum_{n_{x}, n_{y}} \langle n_{x} l_{x} n_{y} l_{y} | \kappa K \rangle 
                       \: | n_{x} l_{x} n_{y} l_{y} : L M \rangle , 
                                          \label{unit_1} \\ 
| n_{x} l_{x} n_{y} l_{y} : L M \rangle  = 
   \sum_{\kappa, K}\:\langle n_{x} l_{x} n_{y} l_{y} | \kappa K \rangle 
           \: | \kappa K l_{x} l_{y} : L M \rangle, \label{unit_2} 
\end{gather} 
are discussed in detail in Refs.~\cite{JiKr,29}. The transformation
coefficients\linebreak 
$\langle n_{x} l_{x} n_{y} l_{y} | \kappa K \rangle$ are
non-zero if only the 
the basis functions  (\ref{11Li-basis}) and (\ref{Jac_fun}) are
characterized by the same number of oscillator quanta $N$,
i.~e. when\linebreak   
${2n_{x} + l_{x} + 2n_{y} + l_{y}} = 2\kappa + K$. 

We suppose that the 1st and the 2nd particles are neutrons and the
3rd particle is the $^9$Li cluster (or the $^4$He cluster in the next
section). Therefore the Jacobi coordinate {\boldmath$x$} is
proportional to the distance between the valent neutrons and  the
Jacobi coordinate {\boldmath$y$} is proportional to the distance
between the cluster and the center of mass of two valent neutrons.
In this case the basis (\ref{Jac_fun}) can be used directly for the
calculation of the $n{-}n$ interaction matrix elements. For the
calculation of the matrix elements of the 
$n{-}{\rm cluster}$ interaction,  we  renumerate the particles assigning
number 1 to the cluster and numbers 2 and 3 to the valent neutrons,
and use Eqs.~(\ref{Jac-x})--(\ref{Jac-y}) to define another set of
Jacobi coordinates {\boldmath$x'$} and {\boldmath$y'$}. The $n{-}{\rm
cluster}$ interaction matrix elements can be directly calculated now
in the basis (\ref{Jac_fun}). The unitary transformation of the wave
functions associated with the switching from one set of Jacobi
coordinates ({\boldmath$x$} and {\boldmath$y$}) to another 
({\boldmath$x'$} and {\boldmath$y'$}), is discussed  in detail in
Refs.~\cite{JiKr,29}. 

\begin{figure}
\centerline{\hspace{-5pt}
\includegraphics[width=0.5\textwidth,angle=0]{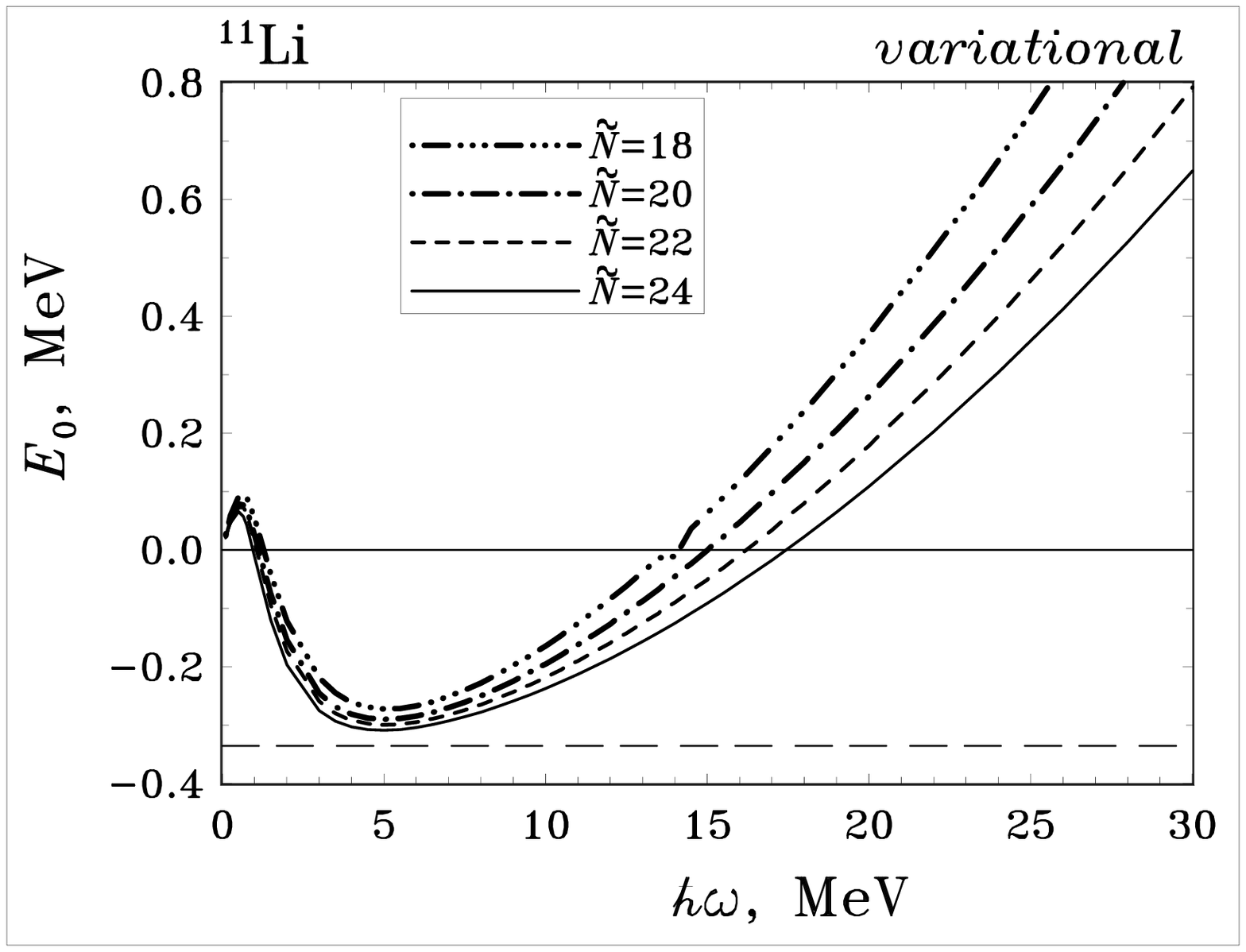}
\hfill
\includegraphics[width=0.5\textwidth,angle=0]{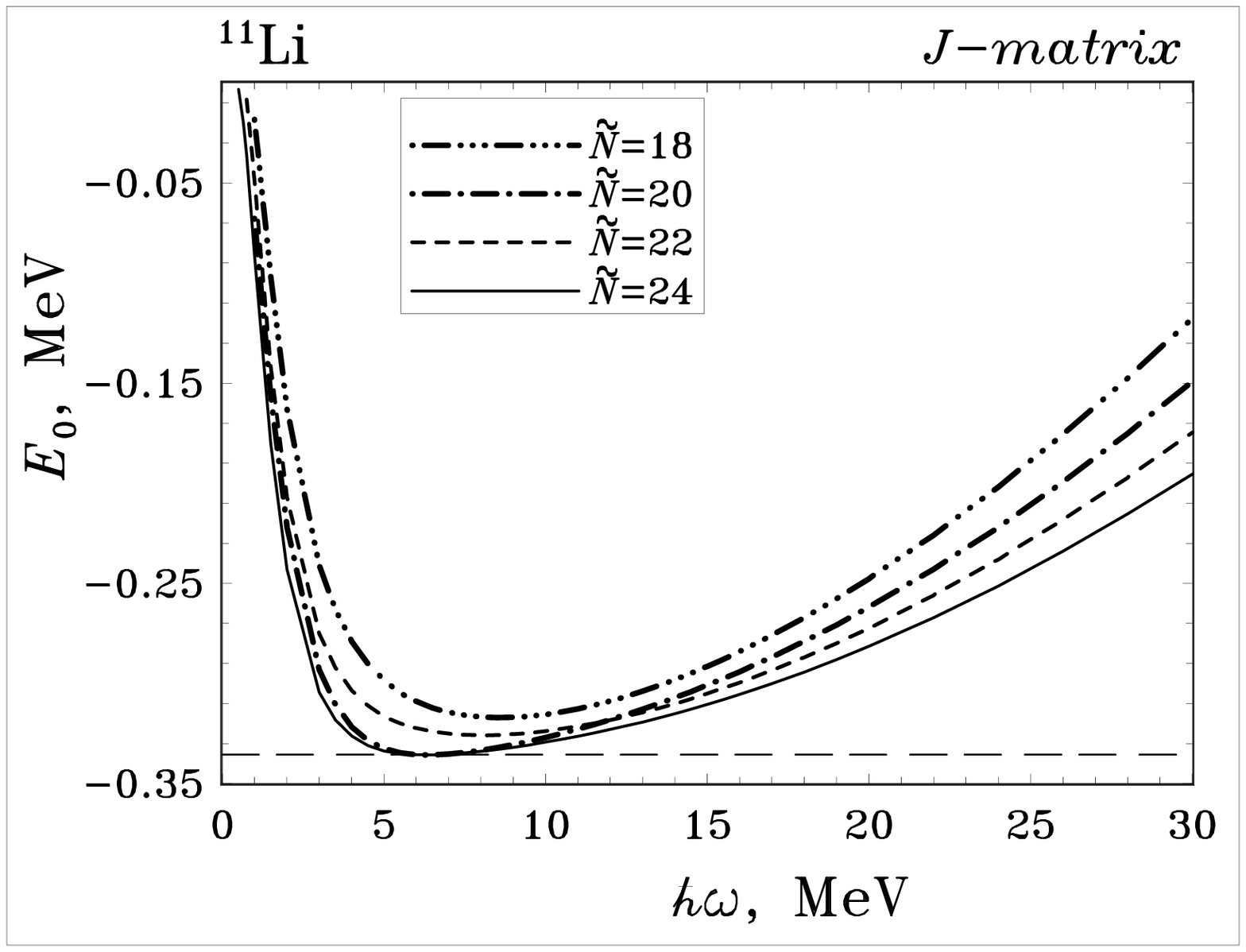}
                   }

\caption{The $^{11}$Li ground state energy vs
the oscillator basis parameter $\hbar\omega$ in the  variational 
(left panel) and $J$-matrix (right panel) calculations for different
values of the truncation boundary $\tilde N$. The horizontal  dashed line
depicts the convergence limit for the ground state energy.
         }
\label{convergence_hw_Li}
\end{figure}

The $^{11}$Li ground state energy dependence on the 
oscillator basis parameter $\hbar\omega$ is shown on
Fig.~\ref{convergence_hw_Li} for different values for the truncation
boundary $\tilde N$. The dependence is typical for variational
calculations and has a minimum at $\hbar \omega \approx 6.5$~MeV (the
best convergence $\hbar\omega$ value). The  $\hbar\omega$  ground state
energy dependence obtained in variational and $J$-matrix calculations
are of the same type. However this dependence is much less pronounced
in the $J$-matrix calculations (note a different scale on the left and
right panels of Fig.~\ref{convergence_hw_Li}). 
Therefore the  $J$-matrix calculation is much less
sensitive to the choice of the  $\hbar\omega$ value. The  $J$-matrix
ground state 
energy results are  better than the variational ones for any  value of
$\hbar\omega$  and any  value of $\tilde N$. 

\begin{figure} 
\centerline{\hspace{-5pt}
\includegraphics[width=0.5\textwidth,angle=0]{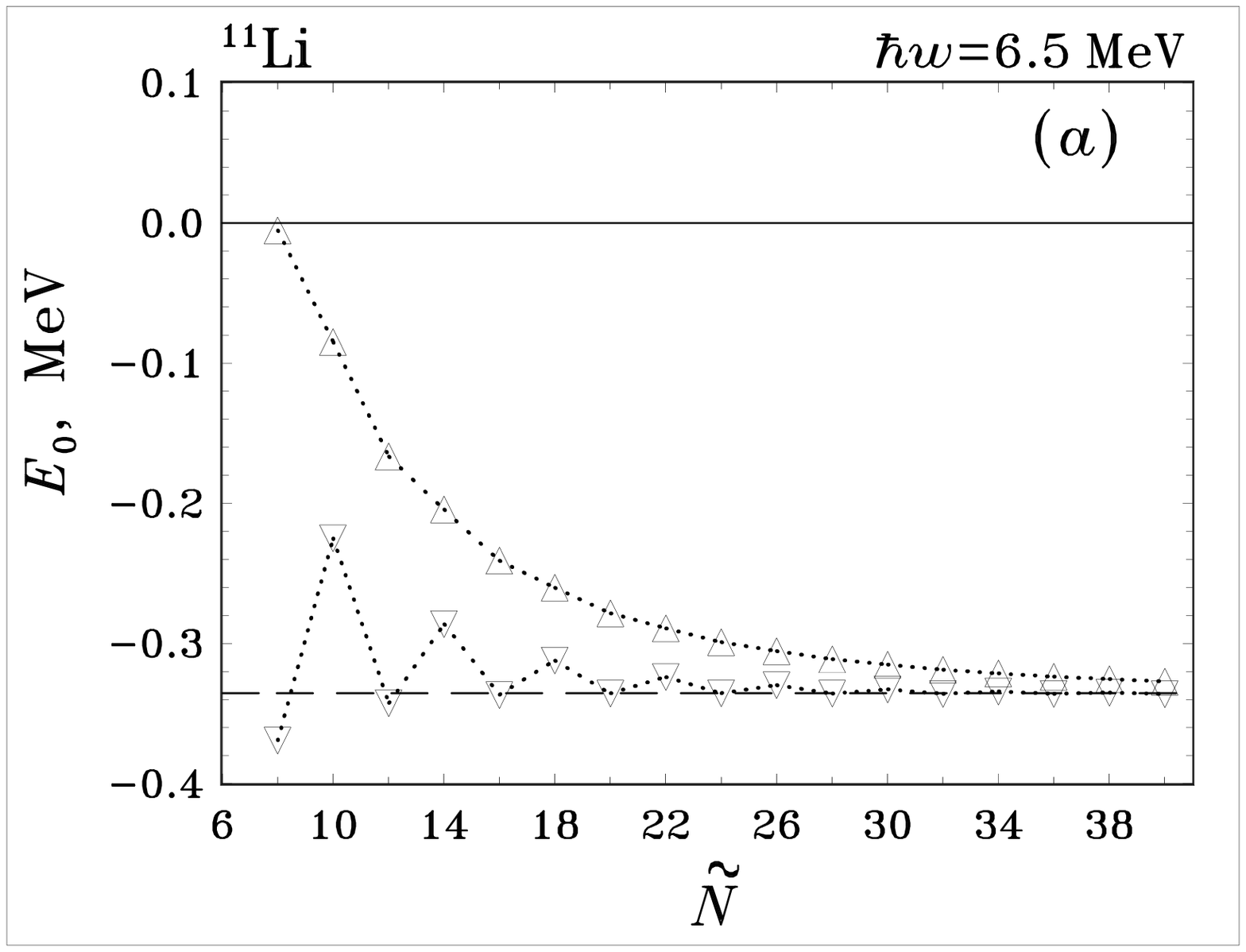}
\hfill
\includegraphics[width=0.5\textwidth,angle=0]{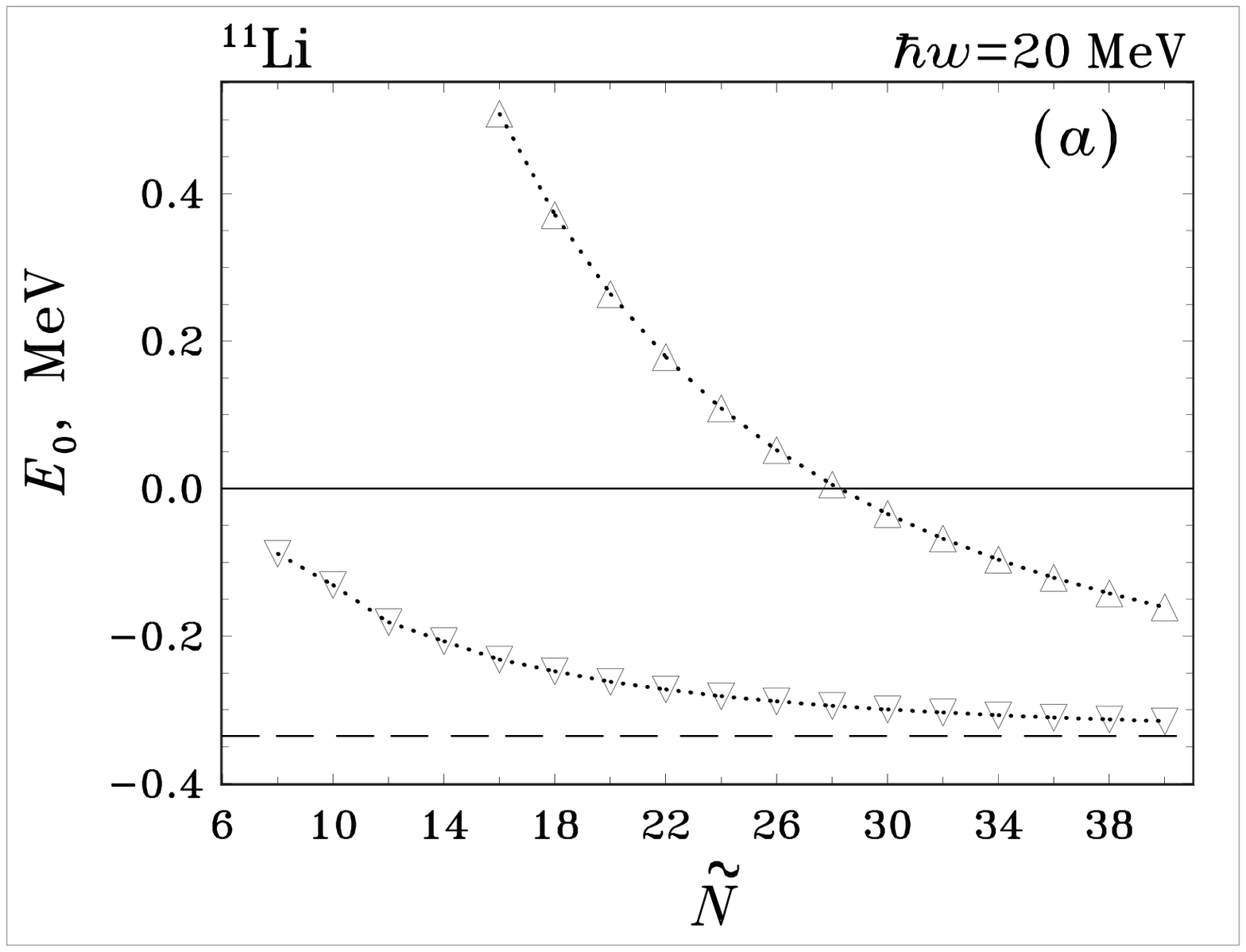}
                   }
\centerline{\hspace{-5pt}
\includegraphics[width=0.5\textwidth,angle=0]{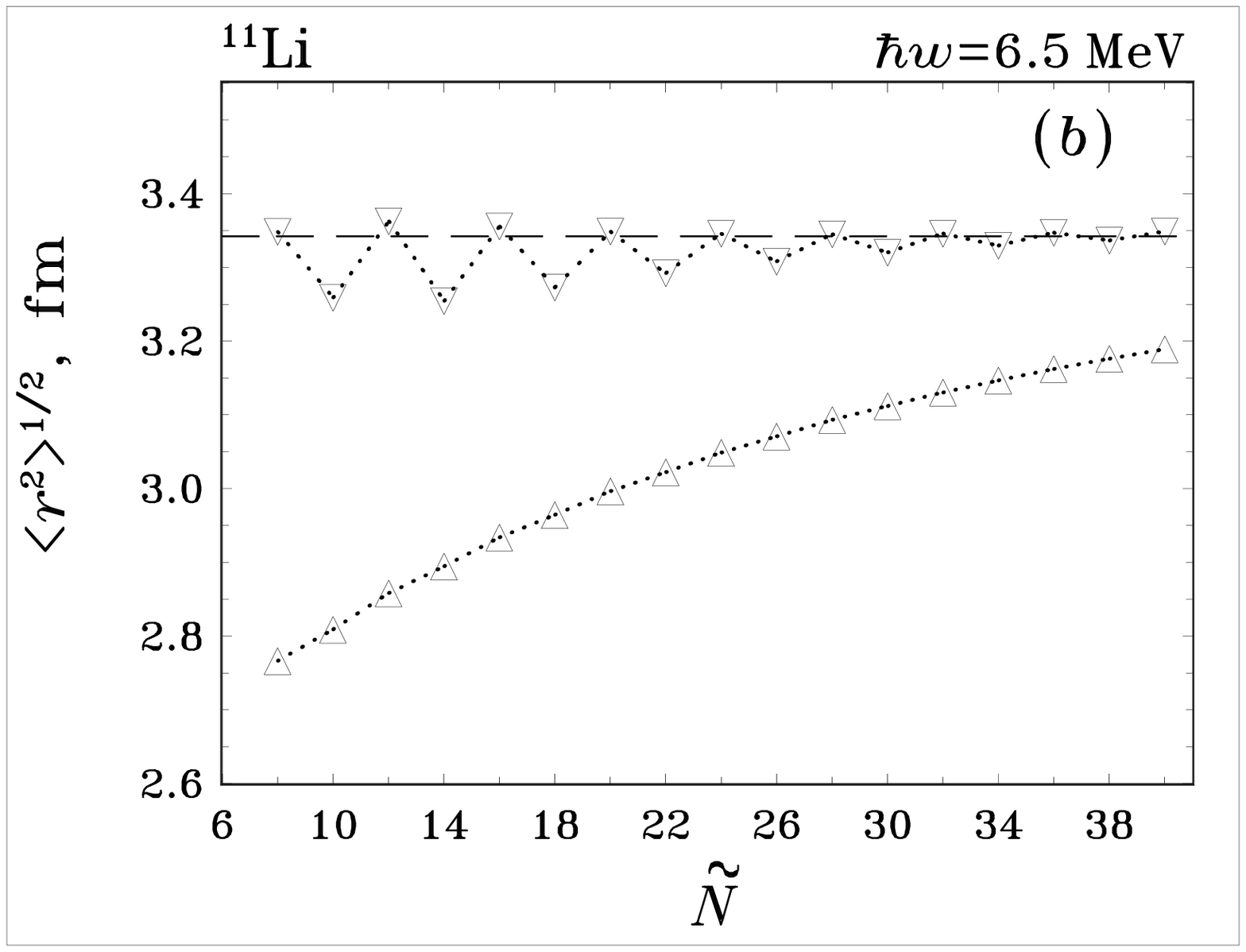}
\hfill
\includegraphics[width=0.5\textwidth,angle=0]{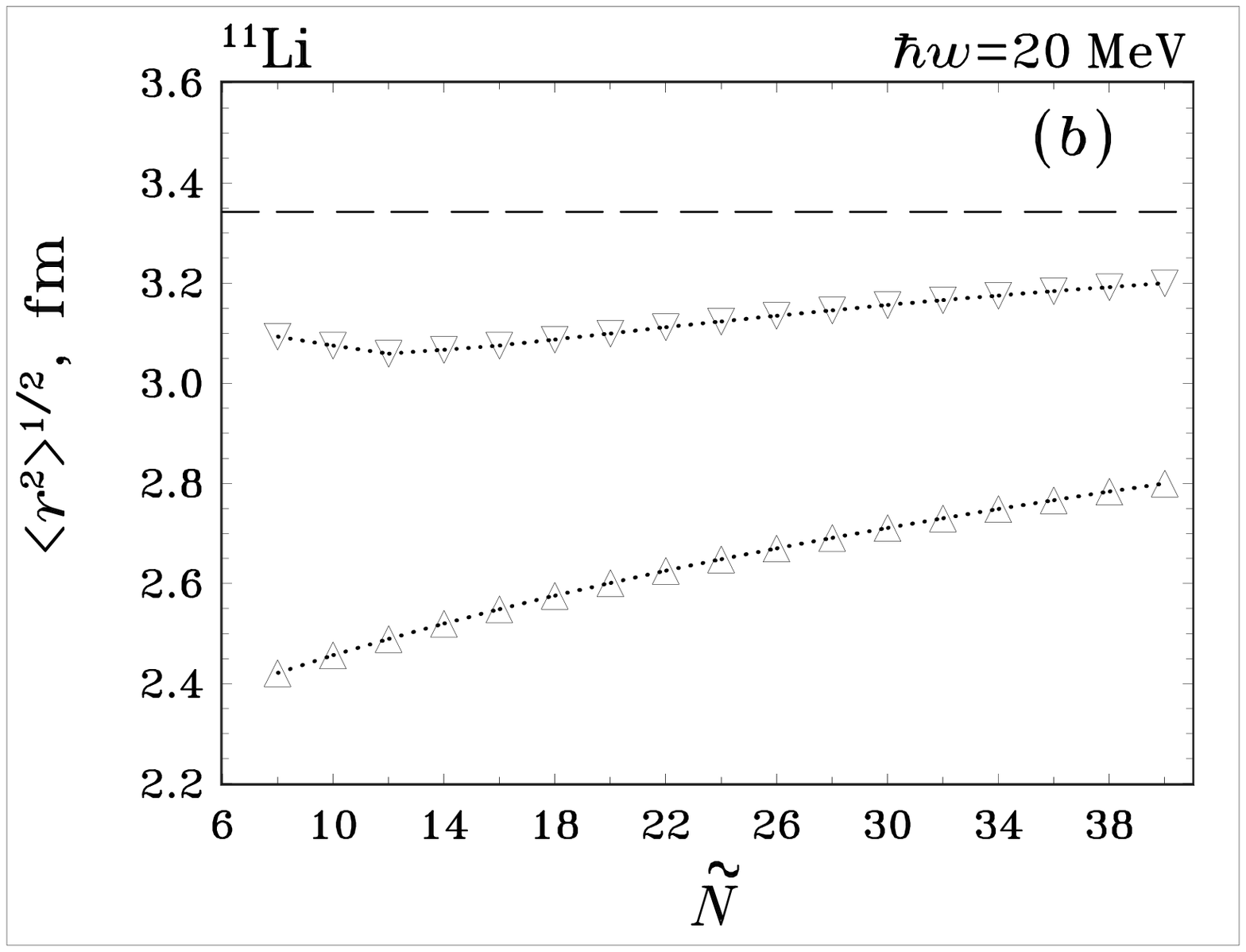}
                   }

\caption{Convergence with $\tilde N$ of the $^{11}$Li three-body
ground state energy (a) and rms matter radius (b)
for $\hbar\omega$=6.5~MeV (left panel) and $\hbar\omega$=20~MeV (right
panel). Variational and $J$-matrix results are shown by
triangles up and triangles down, respectively.  The convergence limits 
are shown by the  dashed line.
}
\label{Li11_convergence_N_hw}

\end{figure}

This is clearly seen on
Fig.~\ref{Li11_convergence_N_hw} where we present the convergence with
$\tilde N$ of the $^{11}$Li ground state energy and rms matter radius 
\begin{equation}
\langle r^2 \rangle ^{1/2} =
\sqrt{
\frac{A-2}{A} \langle r_c^2 \rangle +  \langle \rho^2 \rangle }\,  ,
\end{equation}
where the  $^{11}$Li mass number $A=11$ and
$\langle r_c^2 \rangle^{1/2}$ is the   rms radius of the $^9$Li cluster. 
The variational
ground state energy  decreases  monotonically due to the variational
principal;  the variational rms radius consequently  monotonically
increases. We note here that the variational principle is not
applicable to the $J$-matrix calculations where the infinite
oscillator basis is allowed for and the results obtained with
different $\tilde N$ values differ by the rank of the potential energy
matrix.  The  $J$-matrix results demonstrate a much
more interesting dependence on $\tilde N$ than the variational
ones. For the $\hbar\omega$ 
values close to the best convergence value of $\hbar\omega=6.5$~MeV,
there is a staggering of 
the $J$-matrix binding energy and rms radius with $\tilde N$
(Fig.~\ref{Li11_convergence_N_hw}, left panel). The amplitude of the
staggering decreases with  $\tilde N$ and the results converge
rapidly. It is interesting that the binding energy and the rms radius
converge to the values lying between the results obtained with two
subsequent values of  $\tilde N$. Hence we obtain not only the 
lower bound  binding energy and rms radius estimates as in variational
calculations but both lower bound and upper bound estimates for these
observables. These estimates are presented in
Table~\ref{Li11_energies_radii} together with experimental data and
the results of calculations of other authors.

\begin{table}
\caption{The $^{11}$Li two-neutron separation energy $E(2n)$ and
rms matter radius $\langle r^2 \rangle^{1/2}$ obtained in the calculations
with $\hbar\omega=6.5$~MeV together with the results of theoretical
studies of Ref.~\cite{Danilin-Li11-PhysLett91} and experimental data.
         }
\label{Li11_energies_radii}
\begin{center}
\begin{tabular}{|c|cc|cc|}   \hline
  & \multicolumn{2}{c|}{$E(2n)$, MeV}
            & \multicolumn{2}{c|}{$\langle r^2 \rangle^{1/2}$, fm}   \\
\raisebox{2.ex}[0pt]{Approximation}  & $\tilde N=38$  & $\tilde N=40$
            &  $\tilde N=38$    & $\tilde N=40$  \\ \hline
variational  & 0.326 & 0.327  & 3.176 & 3.189  \\
$J$-matrix    & 0.335 & 0.336  & 3.336 & 3.348  \\
Ref. \cite{Danilin-Li11-PhysLett91}
  &  \multicolumn{2}{c|}{0.3}
       &  \multicolumn{2}{c|}{3.32} \\
\hline 
\raisebox{-1.3ex}[0pt][0pt]{experimental}
  &  \multicolumn{2}{c|}{\raisebox{-1.3ex}[0pt][0pt]{0.247$\pm$0.080, Ref. \cite{Ajzenberg-Selov}}}
    & \multicolumn{2}{l|}{3.10$\pm$0.17,  Ref. \cite{Li11-radii-experiment-old} }  \\
\raisebox{-1.3ex}[0pt][0pt]{data} 
&  \multicolumn{2}{c|}{\raisebox{-1.3ex}[0pt][0pt]{0.295$\pm$0.035, Ref. \cite{Li11-energy-experiment-Benenson}}}
       & \multicolumn{2}{l|}{3.53$\pm$0.10, Ref. \cite{He6-radius} }   \\
  &  \multicolumn{2}{c|}{}
       & \multicolumn{2}{l|}{3.55$\pm$0.10, Ref. \cite{Li11-radii-experiment-Al-Khalili} }   \\   \hline

\end{tabular}
\end{center}
\end{table}

If $\hbar \omega$ differs much from the best convergence value, we
have the conventional variational-type monotonic convergence of the $J$-matrix
ground state energy and rms radius (Fig.~\ref{Li11_convergence_N_hw},
right panel). The results converge slower with $\tilde N$ in this
case, however much faster than in the variational calculation.

\begin{figure}
\centerline{
\includegraphics[width=0.6\textwidth,angle=0]{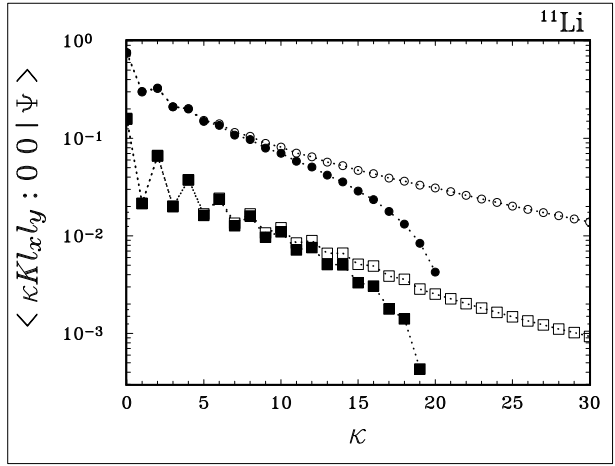}
            }

\caption{
The dominant $K=0$  and $K=2$,  $l_x=l_y=0$  components of the
$^{11}$Li ground state wave function in the     oscillator representation, 
${\langle\kappa,\,K=0,\,l_x=0,\,l_y=0:00\,|\,\Psi\rangle}$ (circles) and  
${\langle\kappa,\,K=2,\,l_x=0,\,l_y=0:00\,|\,\Psi\rangle}$  (squares),
obtained in the 
calculations with $\tilde N=40$ and
$\hbar \omega=6.5$~MeV.
The bold and empty symbols are  the variational
and the $J$-matrix results, respectively.
         }
\label{wf_Li}
\end{figure}

The structure of the ground state wave function is the following. 
The total weight of the  $l_x=l_y=0$ components is 99.15\%, in particular
the $K=0$ component contribution is   94.4\% and the $K=2$,
$l_x=l_y=0$ component 
contribution is  3.3\%. This result agrees well with the results of
other authors who used the three-body cluster model of $^{11}$Li and
the two-body interactions of the same type. In particular,  the
contribution of the  $l_x=l_y=0$ components was estimated in
Ref.~\cite{Hansen-Jensen} as 99\%. According to
Ref.~\cite{Danilin-Li11-PhysLett91},  the weight of the dominant
$K=0$, $l_x=l_y=0$   wave
function component is 95.3\%, and the contribution of this component 
together with the next $K=2$, $l_x=l_y=0$ component  is~98.4\%.
The dominant components of the $^{11}$Li ground state wave function in
the oscillator representation,
${\langle\kappa,\,K=0,\,l_x=0,\,l_y=0:00\,|\,\Psi\rangle}$ and  
${\langle\kappa,\,K=2,\,l_x=0,\,l_y=0:00\,|\,\Psi\rangle}$, are shown in
Fig.~\ref{wf_Li}. It is seen from the figure that the low-$\kappa$
dominating oscillator components are well reproduced in the variational
calculation. However as $\kappa$ approaches the truncation 
boundary~$\kappa_{\Gamma\vphantom{'}}$, the variational calculation
underestimates
$\langle\kappa,\,K=0,\,l_x=0,\,l_y=0:00\,|\,\Psi\rangle$ and 
${\langle\kappa,\,K=2,\,l_x=0,\,l_y=0:00\,|\,\Psi\rangle}$ essentially.

A very interesting problem  is the problem of low-energy electromagnetic
transitions in $^{11}$Li and  other neutron-rich
loosely-bound nuclei. It is supposed that the $E1$ transition strength
is strongly 
enhanced in such nuclei at small enough energies (the so-called {\em
soft dipole mode}). We 
calculate the reduced  $E1$ transition
probability 
\begin{align}
\frac{d\,{\mathcal B} ( E1 )}{d\,E} &=
\frac{1}{2 J_i + 1} \sum\limits_{J_f}
       \big| \langle J_f || {\mrs M}(E1) || J_i \rangle \big|^2
                                     \label{cluster_B_E1}, 
\intertext{where}
{\mrs M}(E1\mu) &= -\, \frac{Zen}{A}\;
\sqrt{ \frac{\hbar}{\omega} \, \frac{2m + M_c}{2 m M_c} }\;  
y\,Y_{1\mu}(\hat{\vecb{y}}),
\label{cluster_operator_E1}
\end{align}
{\boldmath$y$} is the
Jacobi coordinate proportional to the distance
between the $^9$Li cluster and the center of mass of two valent
neutrons, $m$ is the nucleon mass and $M_c$ is the  $^9$Li cluster mass,
$e$ is the electron charge, the number of protons in
$^{11}$Li $Z=3$, 
the number of valent
neutrons $n=2$, $J_i=3/2$  and $J_f$ are the total angular momenta of
the ground (initial) and  final states, respectively. We note here
that within our three-body cluster model of $^{11}$Li we calculate
only the so-called {\em cluster $E1$ transition strength} associated with
the relative motion of two neutrons and the $^9$Li cluster; the total
$E1$  transition strength includes additionally  excitations of
nucleons forming the  $^9$Li cluster that  manifest itself at higher
energies. Since at low energies the only open decay channel is the
three-body one, ${\rm {^{11}Li}\to{^9Li}}+n+n$, the democratic decay
approximation is well justified for the calculation of the final state
wave function.  

\begin{figure}
\centerline{\hspace{-5pt}
\includegraphics[width=0.5\textwidth,angle=0]{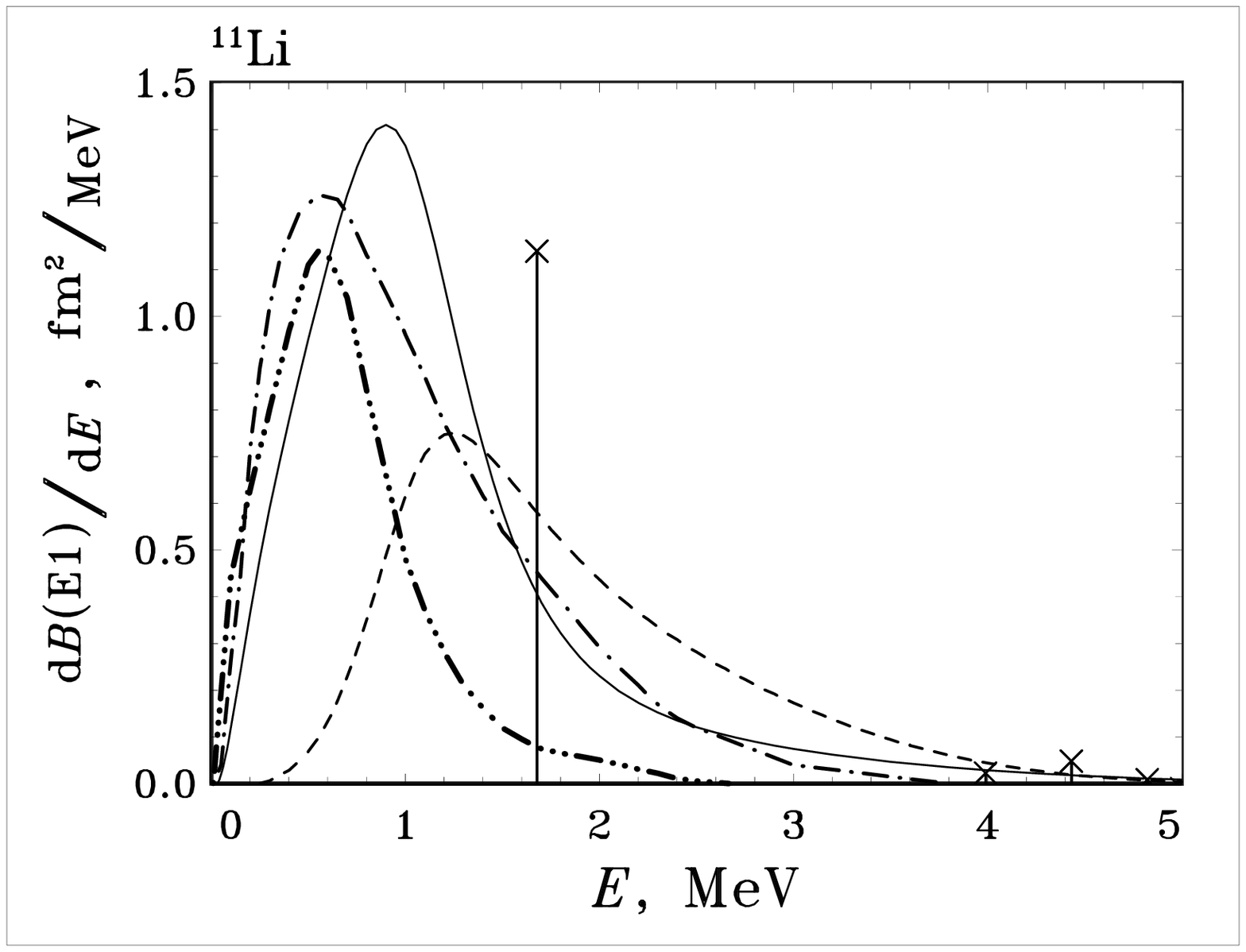}
\hfill
\includegraphics[width=0.5\textwidth,angle=0]{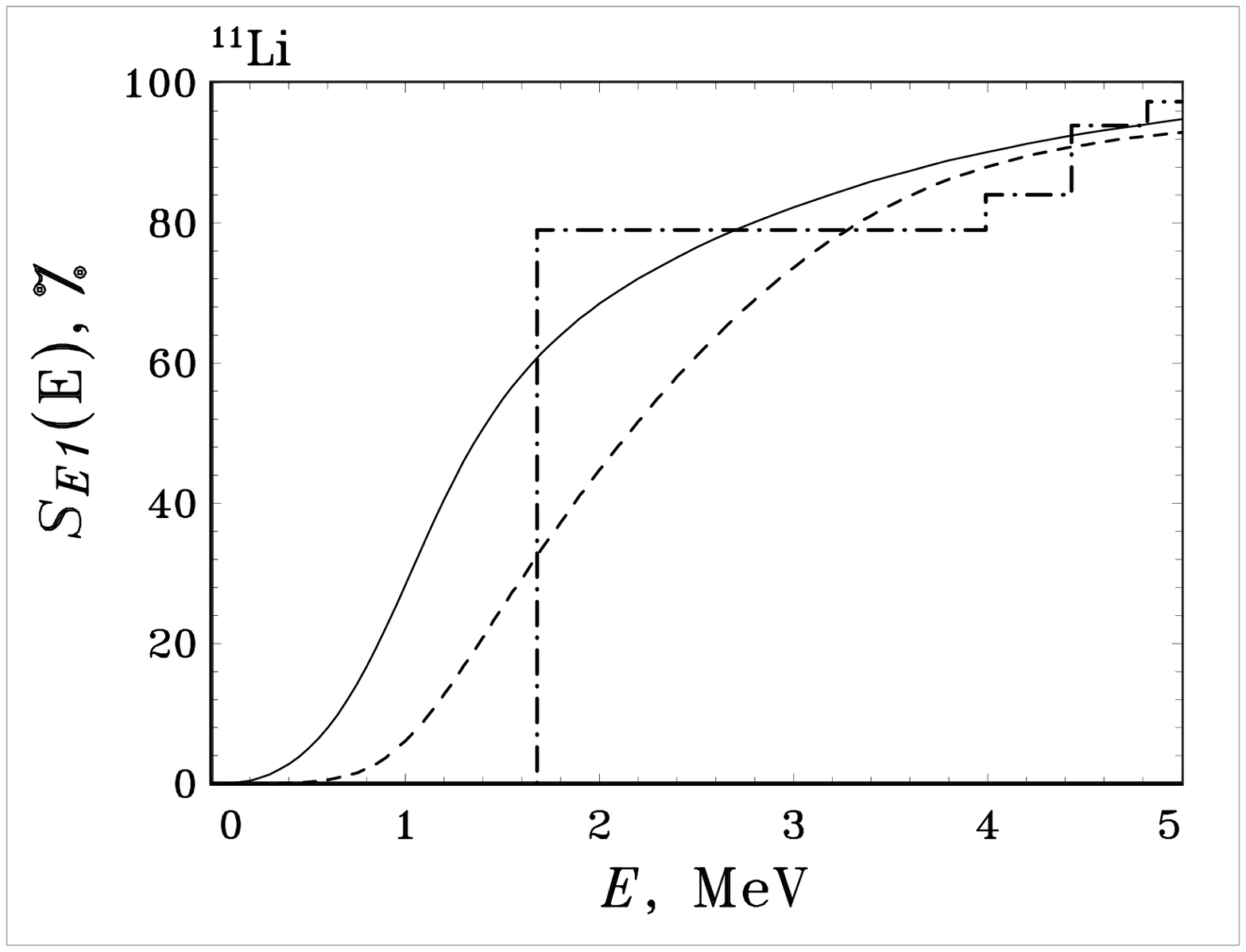}
                   }

\parbox[t]{0.5\textwidth}{
\caption{
Reduced $E1$ transition probability  $\frac{d\,{\mathcal B} ( E1 )}{d\,E}$ 
in $^{11}$Li. Vertical lines with
cross at the end, dashed and solid lines were obtained  in the VV, VJ and JJ
approximations, respectively, with $\hbar\omega=6.5$~MeV and
$\tilde N=20$ for the ground state and $\tilde N=21$ for the final state.
Dash-dot and dash-dot-dot
lines are the calculations of Ref.~\cite{15} and the
parameterizations of experimental data of Ref.~\cite{6}, respectively.
         }
\label{B(E1)_Li11_picture}
}
\hfill
\parbox[t]{0.48\textwidth}{
\caption{
Cluster $E1$ energy-weighted sum rule  in $^{11}$Li.
Dash-dot, dashed and solid lines were obtained in the VV, VJ and JJ
approximations, respectively, with $\hbar\omega=6.5$~MeV and
$\tilde N=20$ for the ground state and $\tilde N=21$ for the final state.
         }
\label{EWSR_E1_Li11_picture}
}
\end{figure}

The results of the 
$\displaystyle\frac{d\,{\mathcal B}(E1)}{d\,E}\vphantom{\int^A}$
calculations are presented in 
Fig.~\ref{B(E1)_Li11_picture}. It is interesting to compare  the
results obtained in the following approximations:
\begin{description}
\item[\rm\hspace{12pt}VV ---] the variational calculation of  both the
ground and 
final  states;
\item[\rm\hspace{12pt}VJ ---] the variational calculation of  the ground
state and the $J$-matrix  calculation of  the final three-body
continuum  state; 

and
\item[\rm\hspace{12pt}JJ ---] the $J$-matrix  calculation of  both the
ground and final three-body  continuum  states.
\end{description}

In the VV approximation, we obtain a discrete spectrum of excited  states and
cannot calculate the final state wave
function at an arbitrary given positive energy. As a result, the $E1$
transition strength differs from zero at the energies belonging to the
discrete spectrum of final states and 
$\displaystyle\frac{d\,{\mathcal B}(E1)}{d\,E}\vphantom{\int^k}$ has a form
of a number of $\delta$-function peaks shown in the figure by vertical
lines with a cross at the end. The first peak at the excitation energy
of approximately 2~MeV exhausts nearly 80\% of the cluster
energy-weighted sum rule \cite{cluster-EWSR-Sagawa}
\begin{equation}
{\mathcal S}^{clust}_{E1} =
\int \limits_0 \limits^{\infty} (E_f - E_0)\,
      \frac{d\,{\mathcal B} ( E1 )}{d\,E} \ d E_f =
\frac{9}{4\pi} \, \frac{\hbar ^2 e^2}{2 m} \, \frac{n Z^2}{A (A-n)}\, ,
\label{EWSR}
\end{equation}
where 
$E_0$ is the ground state energy. This
is clearly seen from  Fig.~\ref{EWSR_E1_Li11_picture} 
where we
present the plot of the function
\begin{equation}
{\mathcal S}_{E1}(E) =
\frac{1}{{\mathcal S}^{clust}_{E1}}
\int \limits_0 \limits^{E} (E_f - E_0)\,
      \frac{d\,{\mathcal B} ( E1 )}{d\,E} \ d E_f .
\label{EWSR_E-dependence}
\end{equation}
The 2~MeV peak should be related with the soft dipole mode in
$^{11}$Li in the VV approach. The energy of the  soft dipole mode obtained in
the VV approximation differs essentially from  the one of the
experimental $E1$ strength 
low-energy maximum. 

In the VJ approximation, the final energy wave function can be
calculated at any given positive energy. Instead of sharp
$\delta$-peaks, we have a smooth 
$\displaystyle\frac{d\,{\mathcal B}(E1)}{d\,E}\vphantom{\int}$ curve with a
maximum shifted to a lower energy that is   closer to the energy of the
experimental $E1$ strength maximum. This is clearly the result of
allowing for the continuum spectrum effects in the final state wave
function. However the external `asymptotic' part of the model space
with $2\kappa+K>\tilde N$ is not allowed for in the ground state wave
function and hence the final  state wave function components with
$2\kappa+K>\tilde N+2$  do not contribute to 
$\displaystyle\frac{d\,{\mathcal B}(E1)}{d\,E}\vphantom{\int}$.

\begin{figure}
\centerline{
\includegraphics[width=0.6\textwidth,angle=0]{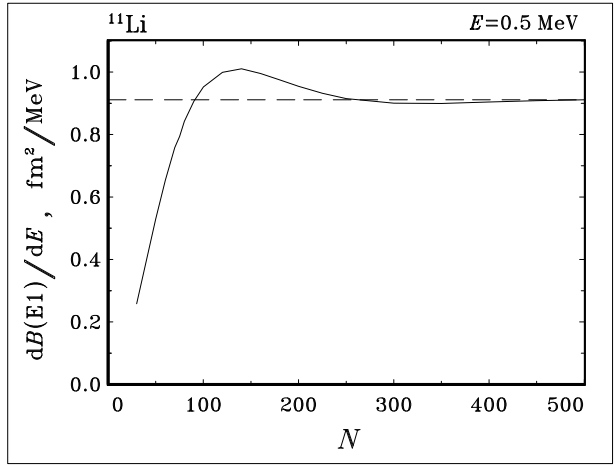}
                   }

\caption{
The $\frac{d\,{\mathcal B} ( E1 )}{d\,E}$
convergence with $N$ at the energy  $E=0.5$~MeV where all
components with the number of oscillator quanta
$2\kappa+K \leq N$ are allowed for in the JJ calculations with
$\hbar\omega=6.5$~MeV and $\tilde N=20$ for the ground state and
$\tilde N=21$ for the final state. 
         }
\label{B(E1)_asymptotic_convergence_Li11}
\end{figure}

The contributions to  
$\displaystyle\frac{d\,{\mathcal B}(E1)}{d\,E}\vphantom{\int}$ of the external
asymptotic part of the model space of both the ground state and the
final state wave functions, are completely accounted for in the JJ
approximation. These contributions are seen from
Figs.~\ref{B(E1)_Li11_picture} and \ref{EWSR_E1_Li11_picture} to be of great
importance. They shift the   
$\displaystyle\frac{d\,{\mathcal B}(E1)}{d\,E}\vphantom{\int}$ maximum to
lower energy and change the shape of the $E1$ strength function. We
note here that due to the slow decrease in the asymptotic region of
the wave function  of the loosely-bound state, it is needed to allow
for a very large number of components in the asymptotic part of the
model space in the $E1$ strength calculations. To calculate 
$\displaystyle\frac{d\,{\mathcal B}(E1)}{d\,E}\vphantom{\int}$ in the
low-energy region with 
high accuracy, we should allow for  all
components with the number of oscillator quanta
$2\kappa+K \leq N$ 
where $N$ is of the order of 1000.
The convergence of 
$\displaystyle\frac{d\,{\mathcal B}(E1)}{d\,E}\vphantom{\int}$ with $N$ at
the energy $E=0.5$~MeV, is shown in
Fig.~\ref{B(E1)_asymptotic_convergence_Li11}.  
We note here that the 
$N\hbar\omega$ oscillator function has the
classical turning point at 
\begin{equation}
\sqrt{ \frac{\hbar}{m \omega} } \rho
\approx \sqrt{ \frac{2N\hbar}{m \omega} }\, .
\label{ClTurn}
\end{equation}
Therefore supposing, say,  $N=500$, we allow 
for the distances up to 
$\displaystyle\sqrt{ \frac{\hbar}{m \omega} } \rho
\approx 80\ \mbox{fm} $
in the 
$\displaystyle\frac{d\,{\mathcal B}(E1)}{d\,E}\vphantom{\int}$ calculation
with $\hbar\omega=6.5$~MeV.

\section{Two-neutron halo in $\bf^{6}He$ nucleus}
 The $^6$He nucleus is studied in the three-body cluster model
${\rm{^6He}}=\alpha+n+n$.  The $^6$He  two-neutron
 separation energy  is 0.976~MeV \cite{He6-energy-experiment}
that is much less than the excitation energy of the $\alpha$ particle
lowest excited state  and the $\alpha$ particle
disintegration threshold.  As it was already noted,  $^6$He  is a Borromean
 nuclei, i.~e. none of the two-body subsystems $\alpha+n$ and
 $n+n$ has a bound state.

Generally our approach in 
the $^6$He case 
is very close to the one 
used in the $^{11}$Li studies discussed above. However, contrary to
the 
$n{-}{\rm{^{11}Li}}$ interaction, the
$n{-}\alpha$ interaction is well-known. Therefore it is reasonable to
use a more microscopically justified approach based on 
available more realistic $n{-}$cluster and $n{-}n$  potentials. In
particular, it is  reasonable to explicitly allow for the spin
variables in our model. We couple the spins of two valent neutrons
into the total spin $S=0$,~1 (the $\alpha$ particle has a zero spin), the
orbital angular momenta $l_x$ and $l_y$ are coupled to the total
orbital angular momentum $L$, and $L$ and $S$ are coupled to the total
angular momentum $J$. Hence the channel index
$\Gamma=\{K,l_x,l_y,L,S,J\}$, and 
we introduce the following  generalization
of the basis functions (\ref{11Li-basis}):
\begin{equation}
     |\kappa K  l_x l_y (L) (S) : J M \rangle  = 
\sum\limits_{M_L, M_S}  \left( L M_L S M_S | J M \right)\,
 | \kappa K l_x l_y : L M_L \rangle \,  | S M_S \rangle ;
\label{6He-wave_function}
\end{equation}
the $^6$He wave function is given by the general formula
(\ref{hyperr-exp}). In Eq.~(\ref{6He-wave_function}), 
$|SM_S\rangle$ is the spin component of the wave function and 
$| \kappa K l_x l_y : L M_L \rangle$ is given by 
Eqs.~(\ref{11Li-basis})--(\ref{hyp_ang_fun}).
We note here that allowing for the triplet ($S=1$) spin states, we
enlarge essentially the number of basis functions with any given
number of oscillator quanta $N=2\kappa+K$. Hence for any given 
$\tilde N$, we have the truncated Hamiltonian matrix of a much larger
rank, and  we are able to perform calculations with smaller $\tilde N$
than in the $^{11}$Li case.

It would be very interesting to perform the $^6$He studies based on
modern so-called {\em realistic} $NN$ potentials derived from the
meson exchange theory and  perfectly
describing $NN$ scattering data and deuteron observables. 
However the oscillator basis matrices of such potentials are extremely
large and cannot be handled in calculations of lightest nuclei; the
realistic potential matrix elements decrease slowly with the number of
oscillator quanta $N$ and truncation of the Hamiltonian matrix results
in the slow convergence of the results. In practical applications
 usually {\em effective} interactions are used in calculations. 
Microscopic {\em ab initio} approaches (see, e.~g., \cite{no-core})
involve {\em realistic effective interactions} and {\em effective
operators} derived from the  realistic $NN$ potentials. Unfortunately
the $J$-matrix formalism (and other formalisms allowing for the
continuum spectrum effects) is not developed still for the case of 
{\em ab initio} models based on realistic effective $NN$
interactions. In this contribution, we are interested in the continuum
spectrum effects and $S$-matrix pole corrections to the binding
energy, and so we use {\em phenomenological effective
potentials} for $n{-}n$ and $n{-}\alpha$ interactions.  A large number
of phenomenological effective  $n{-}n$ and $n{-}\alpha$ potentials is
available. We perform calculations with  various  $n{-}n$ and
$n{-}\alpha$ potentials   since it is not clear from the very
beginning what is the best choice of these potentials;
we would like also to compare our results with the results of other
authors who used different combinations of these interactions.

The following  effective  $n{-}n$  potentials were employed in our
studies. 

The Gaussian potential~(\ref{GaussPotential})  with $V_0=30.93$~MeV
and $R=1.82$~fm in the singlet ($S=0$) state and  $V_0=60.9$~MeV and
$R=1.65$~fm in the triplet ($S=1$) state  suggested in
Ref.~\cite{Gauss-nn-B&J}, is hereafter refered to as Gs. The
potential was fitted to the $s$ wave nucleon-nucleon phase shifts
only; the interaction between the neutrons in the states with relative orbital
angular momentum $l_x>0$, is neglected.

We also make use of a bit different Gaussian potential suggested in
Refs.~\cite{+15,+17} with  $V_0=31.00$~MeV
and $R=1.8$~fm in the singlet state and  $V_0=71.09$~MeV and
$R=1.4984$~fm in the triplet  state.  The singlet component of this
potential was employed in our $^{11}$Li studies. We shall refer to
this potential as G. 
 
A more realistic Minnesota $n{-}n$ potential of Ref.~\cite{Minnesota}
will be refered to as MN.  This Gaussian potential includes central,
spin-orbit and tensor components.

The lowest single particle  $0s_{1/2}$ is occupied in the $\alpha$ 
particle. There are two conventional approaches to the problem of 
the Pauli forbidden  $0s_{1/2}$ state in the $n+\alpha$ system.
The first approach is to add a phenomenological repulsive 
term to the $s$ wave component of the $n{-}\alpha$ potential. This 
phenomenological repulsion excludes the Pauli forbidden state in 
the $n+\alpha$ system and is supposed to simulate the 
Pauli principle effects in more complicated cluster systems. This 
idea was utilized  in the SBB $n{-}\alpha$ potential suggested in 
Ref.~\cite{Danilin-Hyperhermonics}. This is an $l$-dependent Gaussian 
potential that includes the spin-orbit component.

Another approach is to use a deep attractive $n{-}\alpha$ potentials 
that support the Pauli forbidden $0s_{1/2}$ state in the $n+\alpha$ 
system. The Pauli forbidden states should be excluded in the three-body 
cluster system. The conventional method is to supplement the deep 
attractive $n{-}\alpha$ potential by the projecting 
pseudo potential (see, e.~g. \cite{Kukulin})
\begin{equation}
\lambda | 0s_{1/2} \rangle \langle 0s_{1/2} | .
\label{projector}
\end{equation}
If the parameter $\lambda$ is positive and large enough, the projector  
(\ref{projector}) pushes the Pauli forbidden states to very 
large energies and cleans up the ground and low-lying states 
from the Pauli forbidden admixtures. The  eigenfunction of the 
Pauli forbidden state supported by the deep attractive $n{-}\alpha$ 
potential should be used as $ | 0s_{1/2} \rangle$ (the so-called 
eigen-projector \cite{Kukulin}); in this case the 
pseudo potential (\ref{projector}) does not affect the description 
of the scattering data provided by the initial deep attractive
$n{-}\alpha$ potential. 

The  deep attractive Woods-Saxon $n{-}\alpha$ potential suggested in 
Ref.~\cite{WS-Bang-Gignoux} will be hereafter refered to as WS. We 
use the WS potential parameters suggested in Ref.~\cite{He6-radius} 
where the  radius of the potential was increased in order to fit 
 the $^6$He binding energy.

We use also deep attractive  Majorana
splitting potential suggested in Ref.~\cite{Kukulin} (hereafter 
refered to as MS) and  deep attractive $l$-dependent  
Gaussian potential proposed 
in Ref.~\cite{PHT} (hereafter refered to as  GP). These potentials 
improve the description of scattering data in high ($l>1$) partial 
waves.

\begin{figure}
\centerline{\hspace{-5pt}%
\includegraphics[width=0.5\textwidth,angle=0]{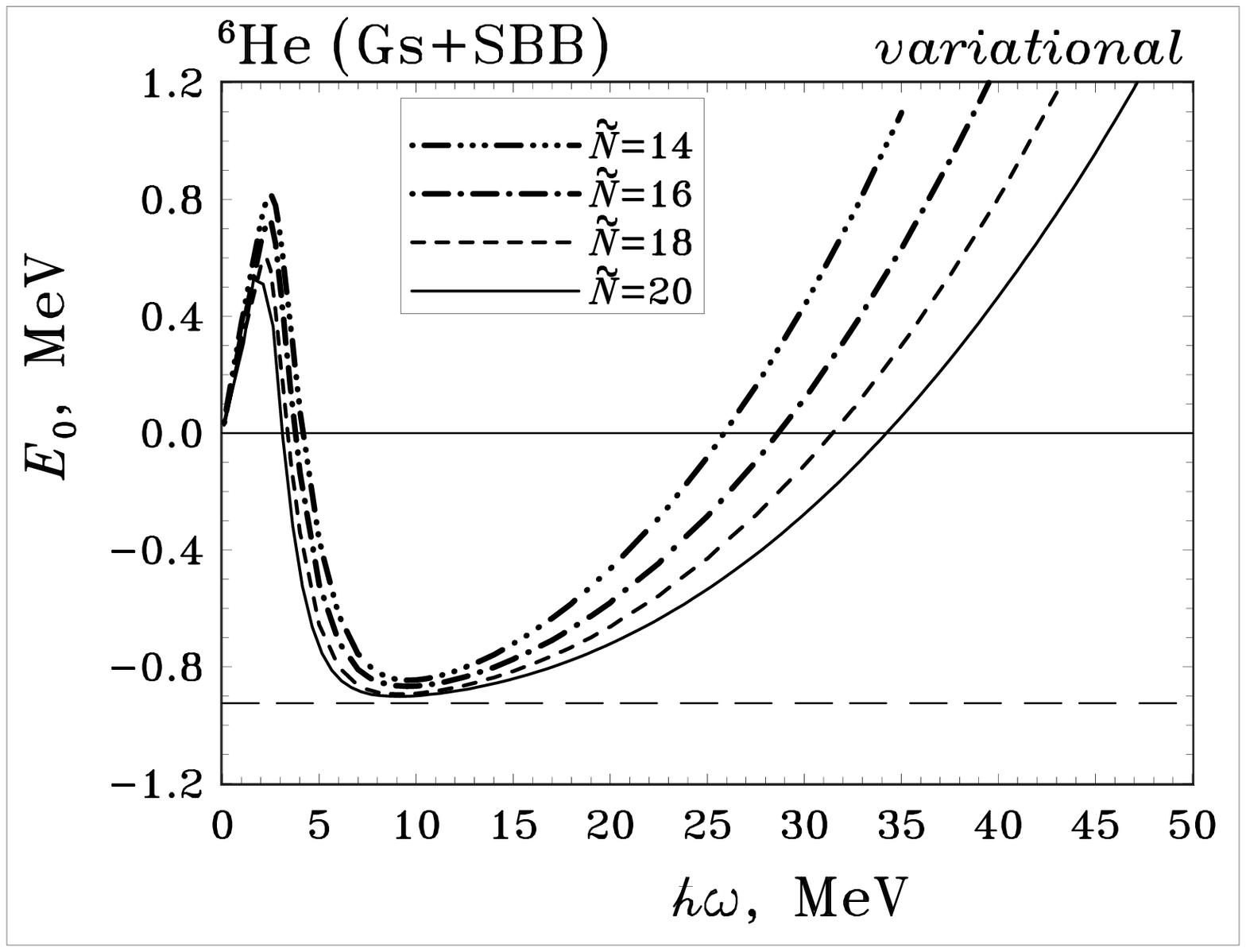}
\hfill
\includegraphics[width=0.5\textwidth,angle=0]{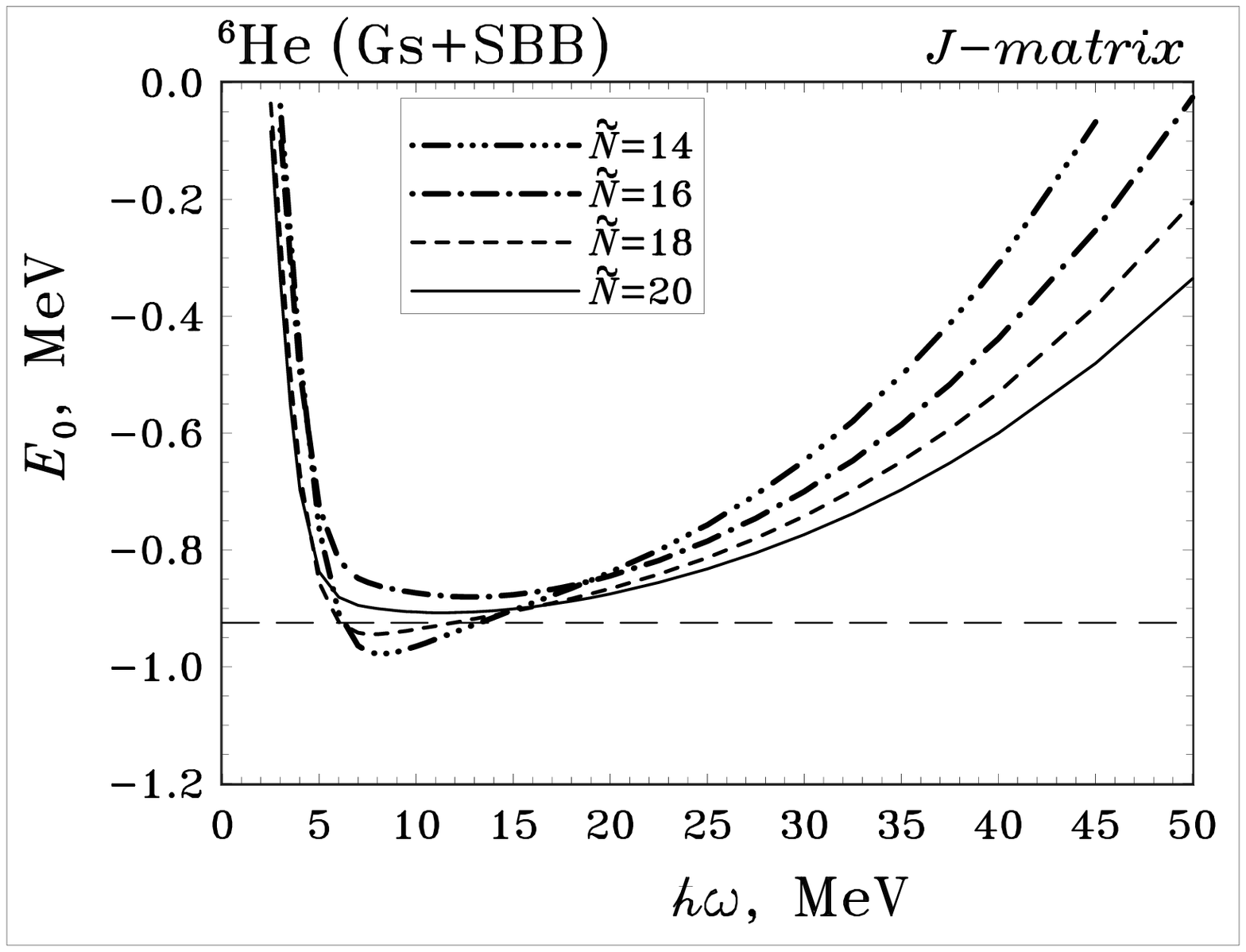}
                   }

\caption{The $^6$He ground state energy in the $\rm Gs+SBB$ 
potential model vs
the oscillator basis parameter $\hbar\omega$ in the  variational 
(left panel) and $J$-matrix (right panel) calculations for different
values of the truncation boundary $\tilde N$. The horizontal  dashed line
depicts the convergence limit for the ground state energy.
         }
\label{convergence_hw_He}
\vspace{8mm}
\centerline{\hspace{-5pt}%
\includegraphics[width=0.5\textwidth,angle=0]{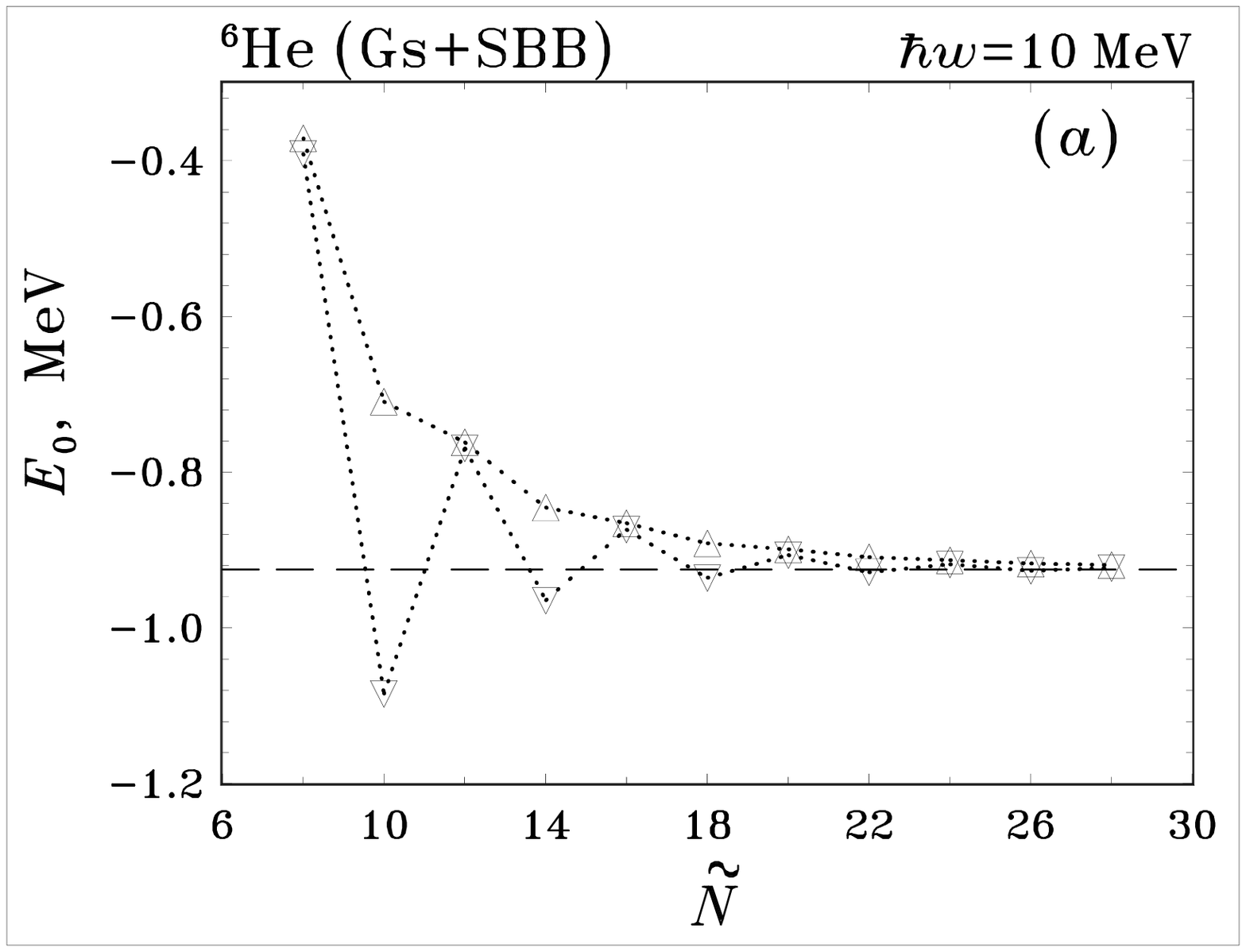}
\hfill
\includegraphics[width=0.5\textwidth,angle=0]{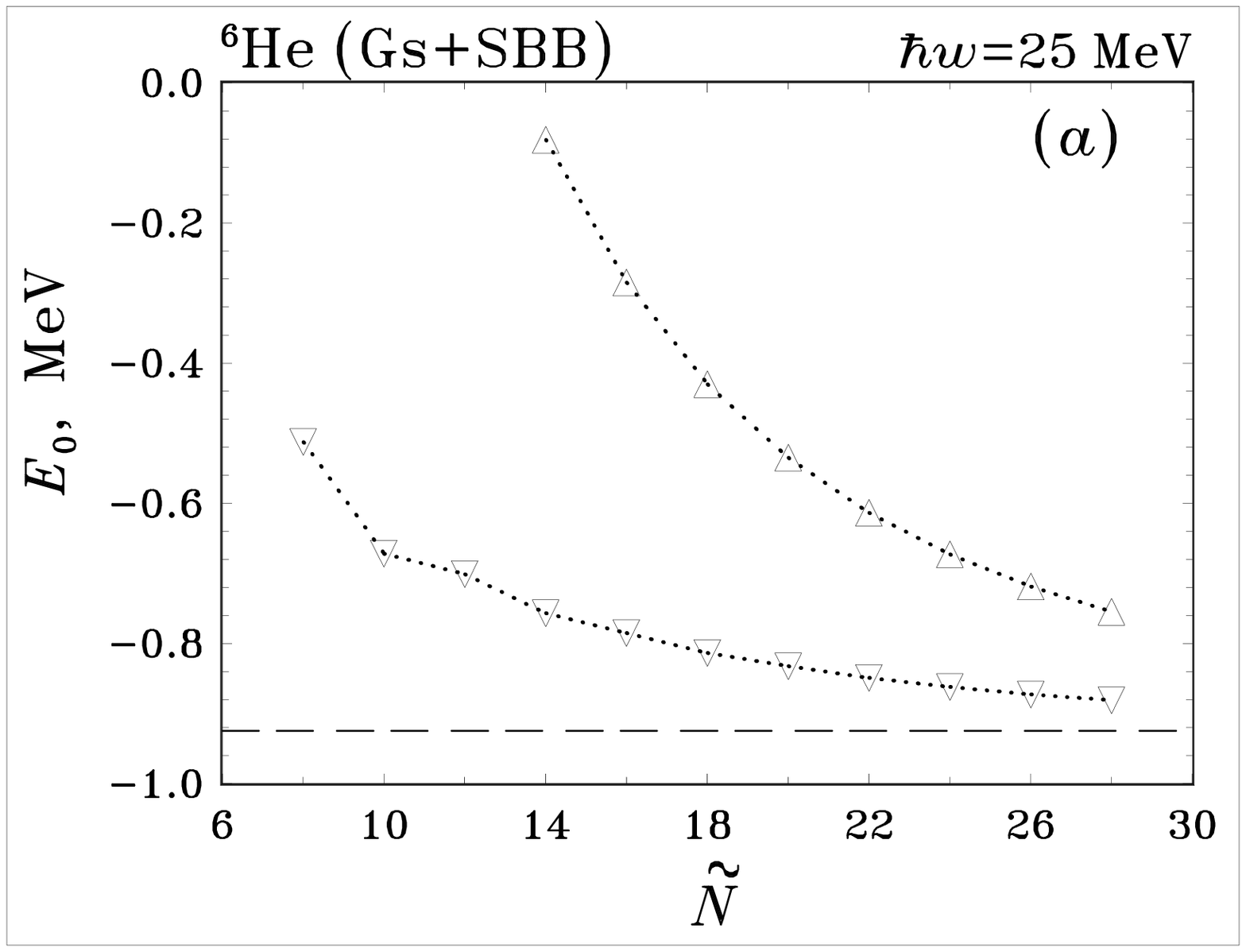}
                   }\vspace{10pt}
\centerline{\hspace{-5pt}%
\includegraphics[width=0.5\textwidth,angle=0]{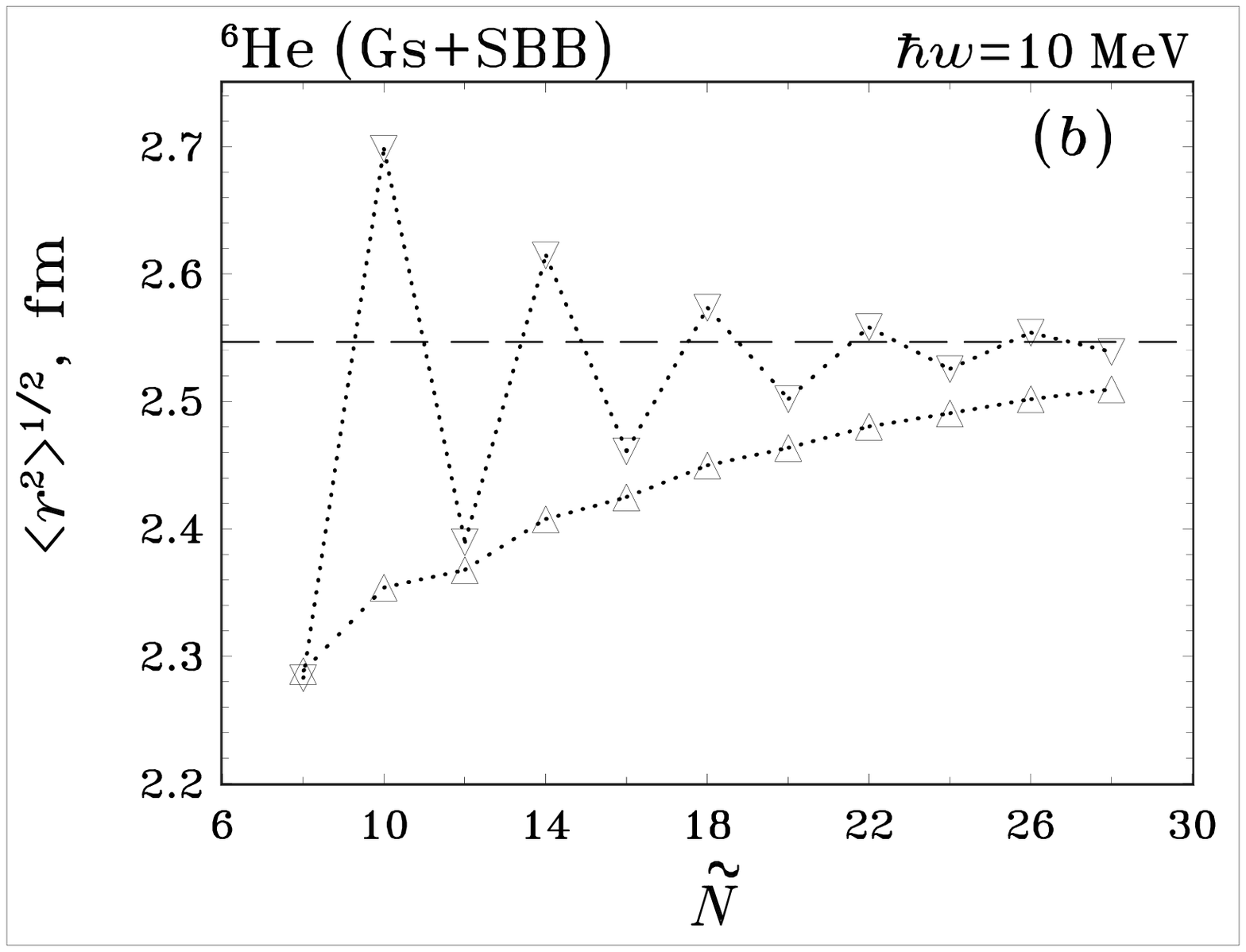}
\hfill
\includegraphics[width=0.5\textwidth,angle=0]{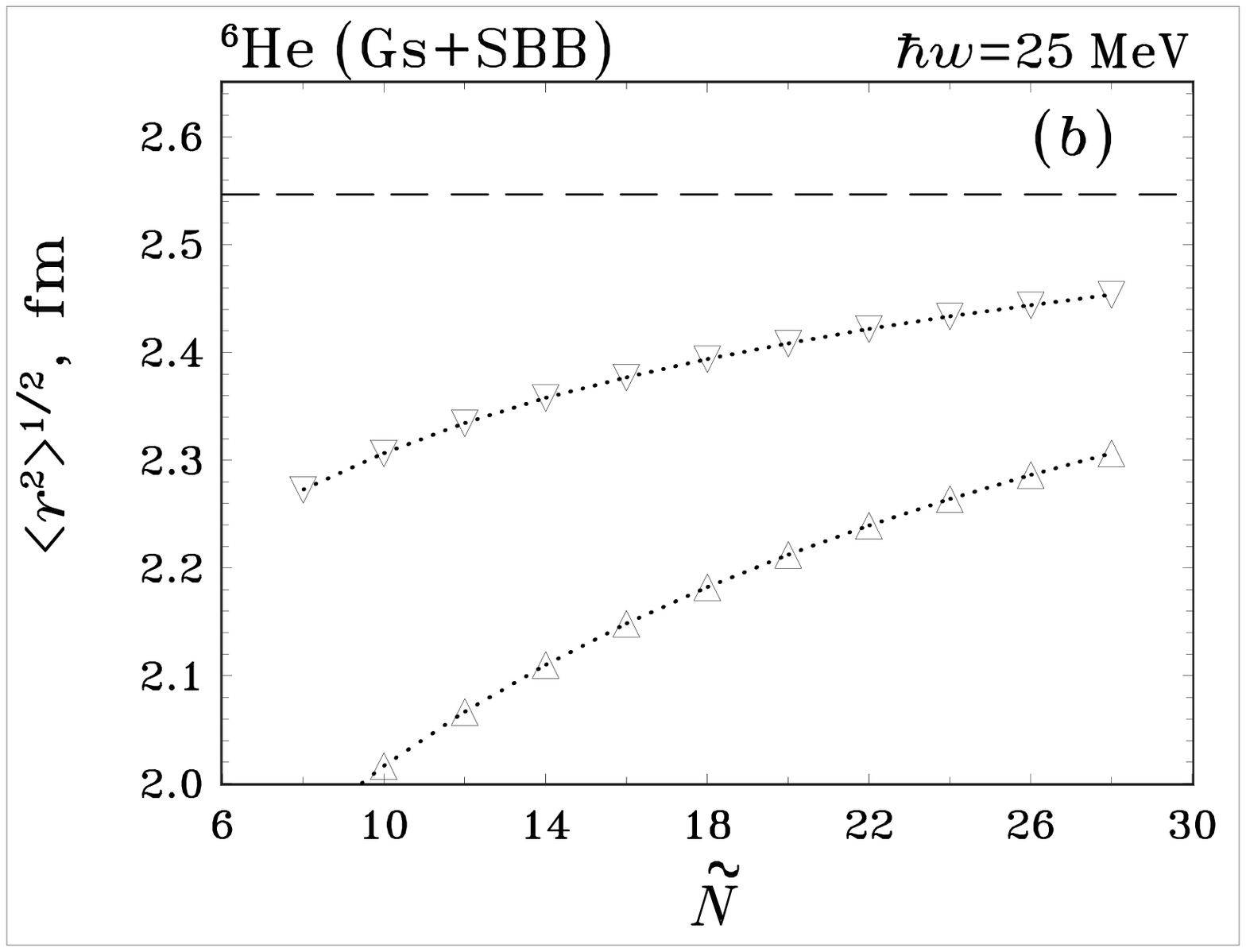}
                   }

\caption{Convergence with $\tilde N$ of the $^6$He ground state energy
(a) and rms matter radius (b) in the $\rm Gs+SBB$ potential model for
$\hbar\omega$=10~MeV (left panel) and $\hbar\omega$=25~MeV 
(right panel). Variational and $J$-matrix results are shown by
triangles up and triangles down, respectively. The convergence limits
are shown by the dashed line.
}
\label{He6_convergence_N_hw}
\end{figure}

The convergence patterns in the $^6$He case are very similar to the ones
discussed  in the $^{11}$Li case. As a typical example, we present the
results obtained in the $\rm Gs+SBB$ potential model.  The $^6$He
ground state energy dependence on 
the oscillator parameter $\hbar\omega$ for a number of truncation
boundary $\tilde N$ values, is depicted in
Fig.~\ref{convergence_hw_He}. The  $\hbar\omega$  ground state energy
dependence is seen to be of the same type as the one in the $^{11}$Li case
shown in Fig.~\ref{convergence_hw_Li}.
The $^6$He ground state energy and rms radius convergence with 
$\tilde N$ is presented in Fig.~\ref{He6_convergence_N_hw} for
two $\hbar\omega$ values. As in the $^{11}$Li case (see
Fig.~\ref{Li11_convergence_N_hw}),  there is the  staggering of the
$J$-matrix binding energy and rms 
radius with $\tilde N$ for $\hbar\omega=10$~MeV that is
close to the value providing the best convergence of the variational
calculation, and there is the variational-type  $\tilde N$ dependence of these
observables for larger  $\hbar\omega$ values. It is interesting that
all calculations shown on  the right panel of
Fig.~\ref{convergence_hw_He}, result in nearly the same binding energy 
at $\hbar\omega\approx 15$~MeV; this  $\hbar\omega$ value is clearly
the best convergence value for the $J$-matrix calculation and it
differs essentially from  the best convergence  $\hbar\omega$ value
for the  variational calculation corresponding to the minima of the
curves on the left panel of Fig.~\ref{convergence_hw_He}.

\begin{table} 
\begin{sideways}
\parbox{\textheight}{
\caption{The $^6$He two-neutron separation energy and rms matter
radius obtained in various potential models and available three-body
cluster model results of other authors.
         }\vspace{5pt}
\label{He6_energies_radii}
\begin{center}
\begin{tabular}{|c|c|c|>{\hspace{.9cm}}cc|>{\hspace{.3cm}}cc|}   \hline
\raisebox{-1.5ex}[0pt][0pt]{Potential model} 
&\raisebox{-1.5ex}[0pt][0pt]{$\hbar\omega$, MeV}&
\raisebox{-1.5ex}[0pt][0pt]{Approximation}  & \multicolumn{2}{c|}{$E(2n)$, MeV}
            & \multicolumn{2}{c|}{$\langle r^2 \rangle^{1/2}$, fm}   \\[-1ex] 
&         &     & $\tilde N=26$  & $\tilde N=28$
            &  $\tilde N=26$    & $\tilde N=28$       \\ \hline
 &\raisebox{-1.5ex}[0pt][0pt]{10.00}& variational& 0.917 & 0.919  & 2.502 & 2.510  \\[-1ex]
\raisebox{-.7ex}[0pt][0pt]{$\rm Gs+SBB$} 
&& $J$-matrix   & 0.927 & 0.923  & 2.554 & 2.539  \\
          & \multicolumn{2}{c|}{\raisebox{0pt}[3ex]{Ref.  \cite{Danilin-Hyperhermonics}\hspace{.5cm}}}
      & \multicolumn{2}{c|}{``{\em correct asymptotic value}''}   &     &
                   \\ \hline 
  &\raisebox{-1.5ex}[0pt][0pt]{11.90}&variational&0.993&0.997&2.394&2.404 \\[-1ex]
\raisebox{-.7ex}[0pt][0pt]{$\rm Gs+WS$} 
&  & $J$-matrix   & 1.003 & 1.008  & 2.444 & 2.461   \\
 &\multicolumn{2}{c|}{\raisebox{0pt}[3ex]{Ref. \cite{He6-radius}\hspace{.5cm}}}
                              & \multicolumn{2}{c|}{1.00}
                      &    \multicolumn{2}{c|}{2.44}   \\ \hline 
\raisebox{-1.5ex}[0pt][0pt]{$\rm G+GP$} 
&\raisebox{-1.5ex}[0pt][0pt]{11.35}
      & variational   & 0.875 & 0.878  & 2.386 & 2.393    \\[-1ex]
&  & $J$-matrix    & 0.889 & 0.883  & 2.451 & 2.428  
                  \\ \hline
\raisebox{-1.5ex}[0pt][0pt]{$\rm MN+GP$} 
&\raisebox{-1.5ex}[0pt][0pt]{10.75}
& variational   & 0.494 & 0.509  & 2.477 & 2.484    \\[-1ex]
&   & $J$-matrix    & 0.515 & 0.516  & 2.588 & 2.547 
                     \\ \hline
\raisebox{-1.5ex}[0pt][0pt]{$\rm MN+MS$}
&\raisebox{-1.5ex}[0pt][0pt]{13.00}
& variational  & 0.656 & 0.672  & 2.382 & 2.389    \\[-1ex]
&  & $J$-matrix   & 0.684 & 0.681  & 2.482 & 2.445
                        \\ \hline 
\multicolumn{3}{|c|}{other potential models}
    & \multicolumn{2}{c}{\parbox{55mm}{Ref. \cite{Kukulin}:  0.305; \   
Ref. \cite{Fedorov-potentiales}:  1.00;\\[.4ex] 
 Ref. \cite{+26}:  0.784; \ Ref. \cite{+27}:  0.696}$\vphantom{\displaystyle\int_{s}^{a}}$   }
   &   \multicolumn{2}{|c|}{Ref. \cite{Fedorov-potentiales}: 2.50} \\ \hline
\multicolumn{3}{|l|}{}
            &  \multicolumn{2}{c}{}
 & \multicolumn{2}{|c|}{Ref. \cite{He6-radii-experiment-Tanihata92}: 2.33$\pm$0.04 }   \\[-1ex]
\multicolumn{3}{|c|}{experimental data}
            &  \multicolumn{2}{c}{ Ref. \cite{He6-energy-experiment}: 0.976}
 & \multicolumn{2}{|c|}{Ref. \cite{He6-radii-experiment-Tanihata87}: 2.48$\pm$0.03 }   \\[-1ex]
\multicolumn{3}{|l|}{}
            &  \multicolumn{2}{c}{}
  &
   \multicolumn{2}{|c|}{Ref. \cite{He6-radii-experiment-Chulkov-Korsheninnikov}: 2.57$\pm$0.10 } \\  \hline
\end{tabular}
\end{center} 
}\end{sideways}
\end{table}

The results of our calculations of the  $^6$He two-neutron separation
energy and rms matter radius, are
summarized in Table~\ref{He6_energies_radii}. Since the calculations
using different potential models were performed with slightly different
$\hbar\omega$ values, we list these values in the 
Table.  If for a given potential model the theoretical
predictions of other authors using other approximations within the
three-body cluster model 
are available, they are presented in the  
corresponding rows of the 
Table, too. The theoretical predictions  within the
three-body cluster model with other potential models and the
experimental data, are presented in the Table in additional rows.

The $J$-matrix approach is seen from the Table to improve the
variational results for both the binding energy and the rms radius
in the case of any potential model. The
improvement is generally more essential in the case of 
smaller binding energy (it is also seen comparing the
$^6$He results in Table~\ref{He6_energies_radii} with the $^{11}$Li
results in Table~\ref{Li11_energies_radii}). The $J$-matrix approach
is also seen to have a faster convergence than the variational one, and
again the improvement of the convergence rate is more essential in the smaller
binding  energy case. Our results are in good correspondence with the
available results of other authors who used the same potential models.

The structure of the $^6$He ground state wave function is
presented in Table~\ref{He6_weights}. We note here that within the
three-body cluster model ${\rm ^6He}=\alpha+n+n$, the $0^+$ ground
state can be obtained by the coupling of  the Jacobi
orbital momenta $l_x$ and $l_y$ to the total orbital
angular momentum $L$  and  subsequent coupling of $L$ with 
the total neutron spin $S$ into the total angular momentum $J=0$, if
only $L=S$ and $l_x=l_y$. 
We arrange the dominant
components in a different manner. In the Table~\ref{He6_weights}\,(a)
we show the total weights  of all components with given $S=L$ and
$l_x=l_y$, i.~e. we sum the contributions of the components 
with different $K$ and $N=2\kappa+K$ values for the given  $S=L$ and
$l_x=l_y$ values.  In the Table~\ref{He6_weights}\,(b) we collect the
total weights  of all components with given $S=L$ and hypermomentum $K$
summing the contributions with different $l_x=l_y$ values. It is also
interesting to calculate the shell model type component weights  which
are characterized by the orbital momenta $l_{n_1\alpha}$ and
$l_{n_2\alpha}$ of individual neutron motion relatively to the $\alpha$
core. The unitary transformation from the hyperspherical basis
(\ref{6He-wave_function}) to the shell model basis can be found in
textbooks (see, e.~g., \cite{Neud,MoSm}). The $^6$He $0^+$ ground
state wave function can be arranged if only
$l_{n_1\alpha}=l_{n_2\alpha}$. The shell model-type component
weights are listed in the  Table~\ref{He6_weights}\,(c).

\begin{table}
\begin{sideways}
\parbox{\textheight}{
\caption{Dominant component weights in  the $^6$He $0^+$ ground state wave
function obtained with various potential models in the $J$-matrix
approximation  with $\tilde N=28$ (the respective
$\hbar\omega$ values can be found in
Table~\ref{He6_energies_radii}) and the results of other authors in
the three-body cluster model.
           }
\label{He6_weights}
\vspace{8mm}

\noindent
(a) Total contribution of 
the components with given  $S=L$ and $L_x=l_y$ (summation over
possible $K$ and $N$ values)\\
{
\begin{center}
\begin{tabular}{|cc|cccc|cc|c|}   \hline
\multicolumn{2}{|c|}{State} &  \multicolumn{7}{c|}{Weight, \%} \\ \hline
\multicolumn{2}{|c|}{} 
  &   \multicolumn{4}{c|}{present work}
  & \multicolumn{2}{c|}{Ref. \cite{Danilin-Hyperhermonics,DaZhu-rep}}
        & Ref. \cite{Kukulin}           \\ 
\raisebox{1.ex}[0pt]{$S=L$\ \mbox{}} & \raisebox{1.ex}[0pt]{$l_x=l_y$}
    & ${\rm Gs+SBB}$ & ${\rm Gs+WS}$ & ${\rm G+GP}$ &$ {\rm MN+MS}$
    & ${\rm Gs+SBB}$ & ${\rm GPT
\footnote{Gogny--Pires--de~Tourreil$n{-}n$ potential \cite{GPT}.}
+SBB}$ 
& ${\rm RSC\footnote{Reid soft-core $n{-}n$ potential \cite{Rsc}.}
            +SBB^*\footnote{SBB  $n{-}\alpha$ potential with modified
parameters (see Ref.~\cite{Kukulin}).}}$     \\  \hline
   & 0 &  82.1  &  83.6  &  80.2  &  76.6  &  83.13\ &  82.87\ &  88.908\  \\
 0 & 2 &  2.46  &  1.90  &  1.64  &  2.07  &  \ 1.77 &  \ 2.31 &  \ 1.035  \\
   & 4 &  0.18  &  0.14  &  0.11  &  0.15  &  \ 0.02 &  \ 0.58 &           \\  \hline
   & 1 &  14.3  &  13.4  &  17.6  &  20.1  &  14.54\ &  13.96\ &  9.692    \\
 1 & 3 &  0.85  &  0.83  &  0.66  &  0.89  &  \ 0.55 &  \ 0.69 &  0.366    \\
   & 5 &  0.09  &  0.09  &  0.06  &  0.09  &         &  \ 0.06  &          \\  \hline
\end{tabular}
\end{center}
}\mbox{\ \ }
\vspace{3mm}

{$^1$Gogny--Pires--de~Tourreil $n{-}n$ potential \cite{GPT}.}

$^2$Reid soft-core $n{-}n$ potential \cite{Rsc}.

$^3$SBB  $n{-}\alpha$ potential with modified
parameters (see Ref.~\cite{Kukulin}).
}\end{sideways}
\end{table}
\addtocounter{table}{-1}
\begin{table}
\begin{sideways}
\parbox{\textheight}{\caption{Prolongation}
\vspace{8mm}
\noindent
(b) Total contribution of the
components with given $K$ and  $S=L$ (summation over
possible  $L_x=l_y$  values)\\ 
\parbox[t]{0.8\textheight}{
\begin{center}
\begin{tabular}{|cc|cccc|cc|}   \hline
\multicolumn{2}{|c|}{State} &   \multicolumn{6}{c|}{Weight, \%} \\ \hline
\multicolumn{2}{|c|}{} 
  &   \multicolumn{4}{c|}{present work}
  & \multicolumn{2}{|c|}{Ref. \cite{Danilin-Hyperhermonics,DaZhu-rep}} \\ 
\raisebox{1.ex}[0pt]{$S=L$\ \mbox{}} &\raisebox{1.ex}[0pt]{$K$}
    & ${\rm Gs+SBB}$ & ${\rm Gs+WS}$ & ${\rm G+GP}$ & ${\rm MN+MS}$
                               & ${\rm Gs+SBB}$ & ${\rm GPT+SBB}$ \\  \hline
   & 0 &  4.28  &  4.06  &  3.99  &  3.93  &  \ 4.41  &  \ 4.68    \\
   & 2 &  77.2  &  78.9  &  75.5  &  72.4  &  78.93\  &  78.10\    \\
 0 & 4 &  0.59  &  0.19  &  0.13  &  0.16  &  \ 0.53  &  \ 0.64    \\
   & 6 &  1.60  &  1.61  &  1.40  &  1.73  &  \ 1.16  &  \ 1.43    \\
   & 8 &  0.67  &  0.51  &  0.43  &  0.41  &  \ 0.39  &            \\  \hline
   & 2 &  14.2  &  13.2  &  17.5  &  20.0  &  13.91\  &  13.41\    \\
   & 4 &  0.12  &  0.15  &  0.05  &  0.06  &  \ 0.12  &  \ 0.15    \\
\raisebox{1.3ex}[0pt]{1}& 6 & 0.82 & 0.80 & 0.63 & 0.85 & \ 0.53 & \ 0.67    \\
   & 8 &  0.05  &  0.04  &  0.04  &  0.06  &  \ 0.02  &            \\  \hline
\end{tabular}
\end{center}
}
\hfill \mbox{}   \\[3mm]
}\end{sideways}
\end{table}
\addtocounter{table}{-1}
\begin{table}
\begin{sideways}
\parbox{\textheight}{\caption{Prolongation}
\vspace{8mm}
\noindent
(c) Total contribution of the
components with given shell model-like orbital angular momenta
$l_{n_1\alpha}=l_{n_2\alpha}$\\
%
\parbox[t]{0.8\textheight}{
\begin{center}
\begin{tabular}{|c|cccc|c|}   \hline
 State  &   \multicolumn{5}{c|}{Weight, \%} \\ \hline 
   &   \multicolumn{4}{c|}{present work}
            &Ref. \cite{Thompson-Danilin-PauliBlocking} \\ 
\raisebox{1.3ex}[0pt]{$l_{n_1\alpha}=l_{n_2\alpha}$}
    & ${\rm Gs+SBB}$ & ${\rm Gs+WS}$ & ${\rm G+GP}$ & ${\rm MN+MS}$
& ${\rm GPT+WS+V_3\footnote{Three-body $nn\alpha$ potential, see 
Ref.~\cite{Thompson-Danilin-PauliBlocking}.}}$ \\  \hline
$s$  &  8.27  &  7.60  &  7.35  &  7.26  & \ 7.7  \\
$p$  &  90.4  &  90.9  &  91.5  &  91.6  &  91.0\ \\
$d$  &  0.57  &  0.64  &  0.43  &  0.44  &        \\
$f$  &  0.57  &  0.54  &  0.47  &  0.41  &        \\
$g$  &  0.07  &  0.07  &  0.07  &  0.05  &        \\     \hline
\end{tabular}
\end{center}
}
\hfill \mbox{}\\[3mm]

$^4$Three-body $nn\alpha$ potential, see 
Ref.~\cite{Thompson-Danilin-PauliBlocking}.
}\end{sideways}

\end{table}

It is seen from  Table~\ref{He6_weights} that different potential
models result in the wave functions of nearly the same
structure. For example, the component weights in different
arrangements obtained with SBB potential
simulating the Pauli principle by repulsive terms and with WS potential
when the Pauli forbidden states are explicitly projected out, are
nearly the same. Our results for  component weights are in good
correspondence with the results of other authors who used other
approaches to the three-body cluster model and other potential
models. Therefore the uncertainties of the $^6$He wave function
structure due to the uncertainties of the two-body interactions, are
very small. The most essential (however not very large) difference
between the dominant component weights obtained in different
approaches  is the difference between the results of
Ref.~\cite{Kukulin} where the $\rm RSC+SBB^*$ potential model was
employed and our results and the results of 
Ref.~\cite{Danilin-Hyperhermonics,DaZhu-rep} obtained with different
potential models. We note here that 
the $^6$He binding energy obtained in  Ref.~\cite{Kukulin} is
0.3046~MeV only, i.~e. it was 
essentially underestimated. Most probably this is indication that the  
$\rm RSC+SBB^*$ potential model is not adequate for the description of
$^6$He in the cluster model and hence it is not surprising that the
component weights of  Ref.~\cite{Kukulin} differ from the ones
obtained with other potentials.

From the naive shell model considerations it follows that two valent neurons
should occupy $p$ states in the $^6$He nucleus. We see from  the
Table~\ref{He6_weights}\,(c) that this situation is really utilized in 
the dominant component of the $^6$He wave function. The same naive
shell model considerations bring us to the $K=2$ component dominance,
and again we see from  the  Table~\ref{He6_weights}\,(b) that this is
really the case. However from these naive  considerations it looks
strange that the dominant component corresponds to the Jacobi orbital
momenta $l_x=l_y=0$ [see the
Table~\ref{He6_weights}\,(a)]. Nevertheless there is, of course, no
contradiction between the results presented in the
Table~\ref{He6_weights}\,(a) and the Table~\ref{He6_weights}\,(c),
they are obtained by the summation over different quantum numbers of
the same $^6$He ground state wave functions.

\begin{figure}
\centerline{
\includegraphics[width=0.6\textwidth,angle=0]{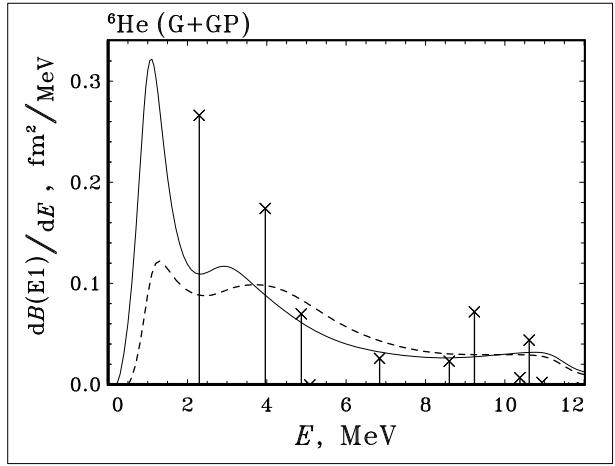}
                   }

\caption{
Reduced $E1$ transition probability  $\frac{d\,{\mathcal B} ( E1 )}{d\,E}$ 
in  $^{6}$He obtained in the $\rm G+GP$ potential model.
Vertical lines with
cross at the end, dashed and solid lines were obtained  in the VV, VJ and JJ
approximations, respectively, with $\hbar\omega=11.35$~MeV and
$\tilde N=22$ for the ground state and $\tilde N=23$ for the final state.
         }
\label{B(E1)_He6_picture}
 \end{figure}

An example of the reduced (cluster) $E1$ transition probability
calculations in $^6$He is shown in 
Fig.~\ref{B(E1)_He6_picture} where the results obtained with 
$\rm G+GP$ potential model  in the VV, VJ and JJ approximations are
presented.  In the $^6$He case, the transition
strength is distributed  in the VV approximation over a number of
strong $\delta$-peaks. Therefore the shape of the reduced probability 
 $\displaystyle\frac{d\,{\mathcal B} ( E1 )}{d\,E}$ appears to be more complicated
than in the $^{11}$Li case when the effects of continuum are allowed
for in the VJ and JJ approximations. As in the $^{11}$Li case, we see
that the external asymptotic part of the model space allowed for in
the JJ approximation only,  provides a very significant contribution to
the electromagnetic transition probabilities.

\begin{figure}
\centerline{\hspace{-5pt}%
\includegraphics[width=0.5\textwidth,angle=0]{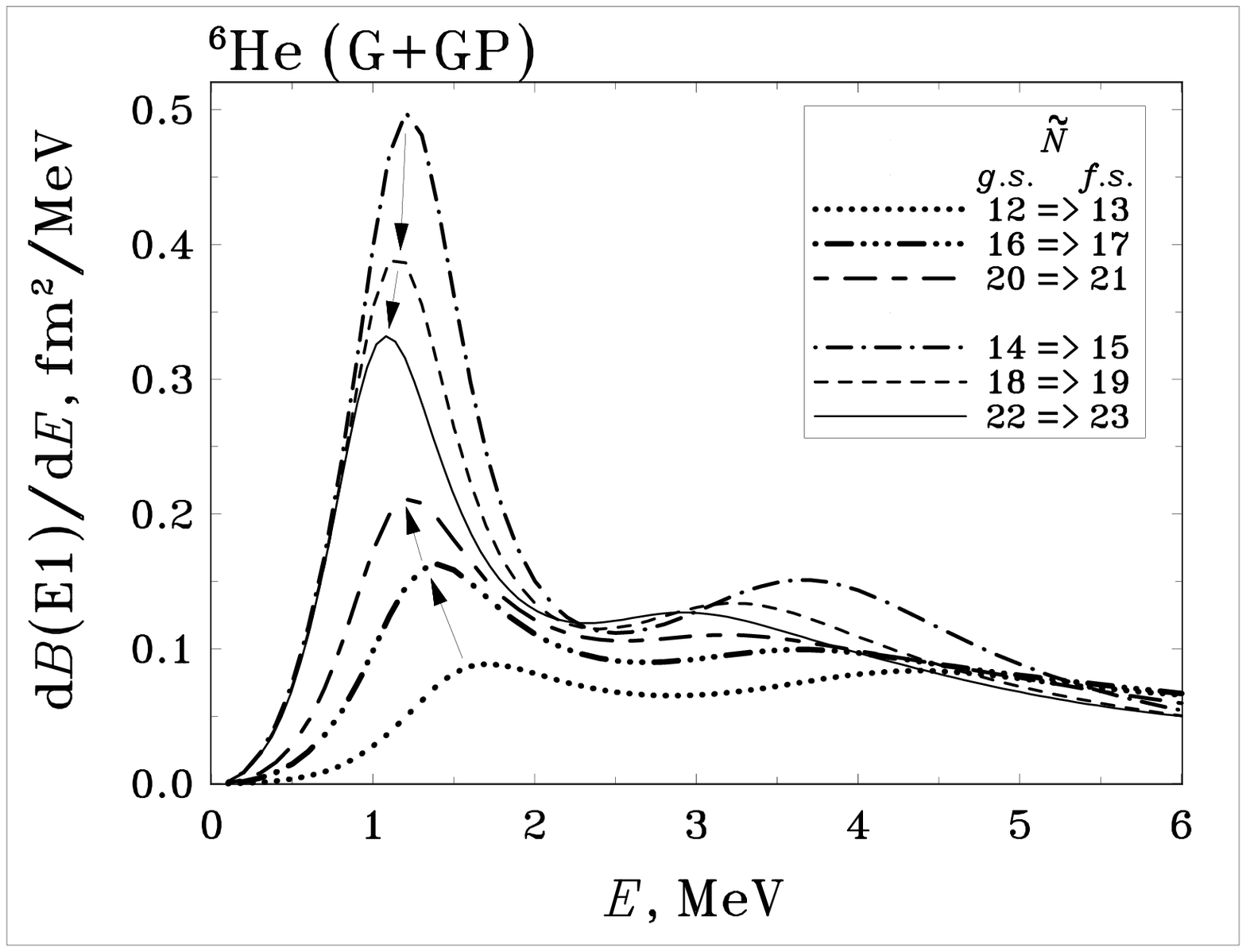}\hfill
\includegraphics[width=0.5\textwidth,angle=0]{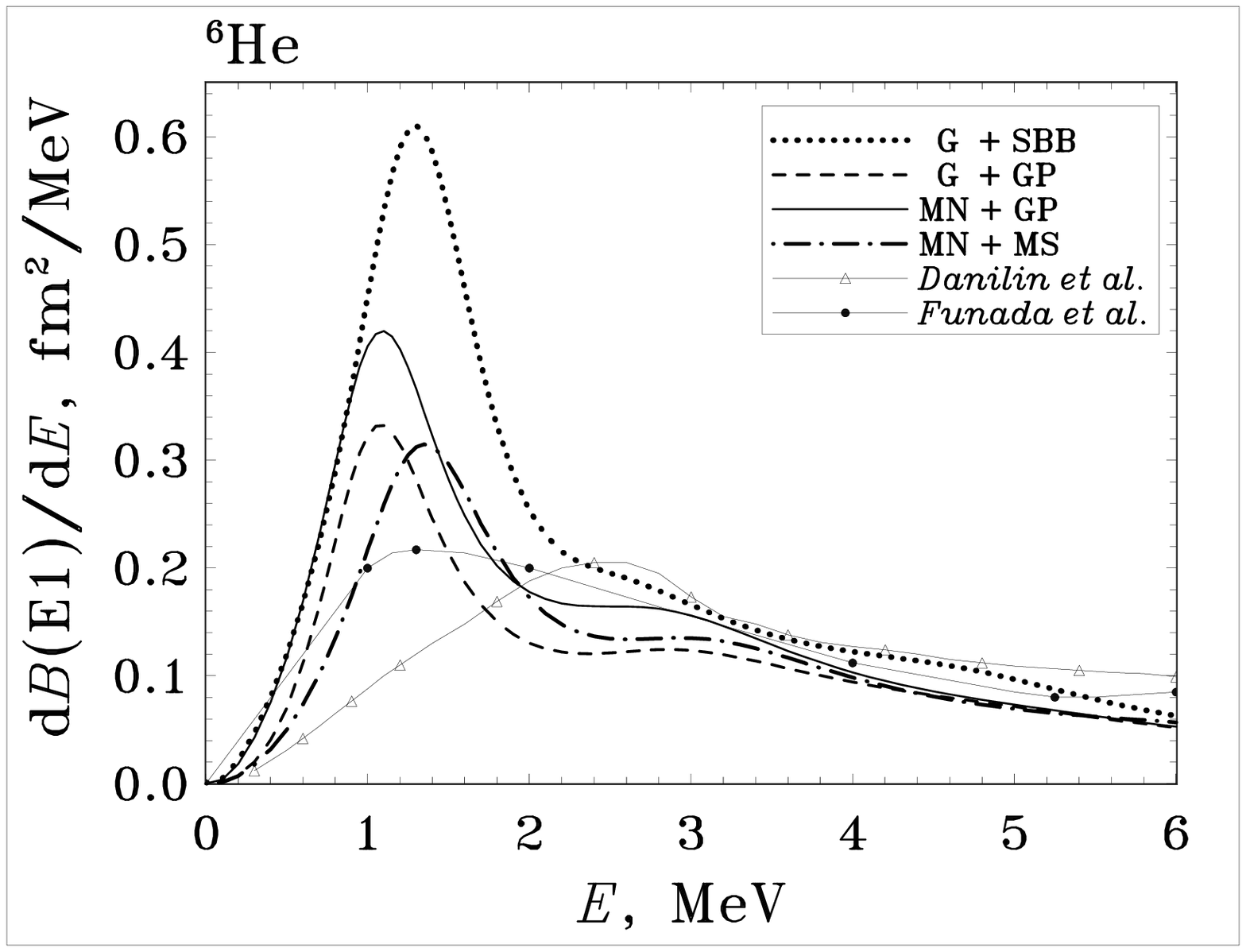}
                   }
\parbox[t]{0.48\textwidth}{
\caption{
Convergence of the $\frac{d\,{\mathcal B} ( E1 )}{d\,E}$ calculations in $^6$He
with the truncation parameter 
$\tilde{N}_{\rm g.s.}$ used in the ground state calculations (the final state 
truncation parameter   $\tilde N_{\rm f.s.}=\tilde N_{\rm g.s.}+1$). 
The results are obtained in the JJ approximation with the
$\rm G+GP$ potential model and  $\hbar\omega=11.35$~MeV. Arrows show
the changes of  $\frac{d\,{\mathcal B} ( E1 )}{d\,E}$ when 
$\tilde{N}_{\rm g.s.}$  is increased by 4.
        }
\label{B(E1)_He_N_covergency}
}
%
\hfill
\parbox[t]{0.48\textwidth}{
\caption{Reduced $E1$ transition probability  
$\frac{d\,{\mathcal B}(E1)}{d\,E}$ in  $^{6}$He obtained  with various
potential models in the JJ approximation with the ground state 
truncation parameter $\tilde{N}_{\rm g.s.}=22$ and the final   state 
truncation parameter $\tilde{N}_{\rm f.s.}=23$ (the corresponding  
$\hbar \omega$ values can be found in  Table~\ref{He6_energies_radii})
in comparison with the results of Funada et al.  \cite{Funada} and 
Danilin et al. \cite{Danilin-PhysRev97}.
         }
\label{B(E1)_He6_comparison}
}
\end{figure}

The convergence of the reduced  $E1$ transition probability
calculations in the JJ approximation is illustrated by
Fig.~\ref{B(E1)_He_N_covergency} where we show the results obtained
with different truncation boundaries in the ground state calculations
$\tilde N_{\rm g.s.}$, the truncation boundary in the final state
calculations $\tilde N_{\rm f.s.}=\tilde N_{\rm g.s.}+1$. The
staggering of the  $\displaystyle\frac{d\,{\mathcal B} ( E1 )}{d\,E}$
shape as  $\tilde N_{\rm g.s.}$ increases, is clearly seen in the
figure. For example, the weakest transition strength is obtained with 
$\tilde N_{\rm g.s.}=12$ and $\tilde N_{\rm f.s.}=13$ shown by the
dotted line in the figure; the strongest transition strength is obtained with 
$\tilde N_{\rm g.s.}=14$ and $\tilde N_{\rm f.s.}=15$ shown by the
dash-dot line; the transition strength   obtained with 
$\tilde N_{\rm g.s.}=16$ and $\tilde N_{\rm f.s.}=17$ is stronger than
the one   obtained with 
$\tilde N_{\rm g.s.}=12$ and $\tilde N_{\rm f.s.}=13$ but weaker than
all the rest results presented in the figure, etc. However selecting
the results obtained with even $\tilde N_{\rm g.s.}/2$ values only, we see that
the $E1$ transition strength increases monotonically with  
$\tilde N_{\rm g.s.}$;   selecting
the results obtained with odd $\tilde N_{\rm g.s.}/2$ values only, we
see the monotonic decrease of the $E1$ transition strength with  
$\tilde N_{\rm g.s.}$.

The comparison of the  reduced  $E1$ transition probability results
obtained with different potential models and calculations of other
authors within the three-body cluster model, is shown in
Fig.~\ref{B(E1)_He6_comparison}.

\section{Phase equivalent transformation with continuous parameters
and three-body cluster system}
\label{PET}

The results presented  in Table~\ref{He6_energies_radii} show that we
obtain a very good approximation for the $^6$He two-neutron separation
energy. However the exact value of the $^6$He binding energy is not
reproduced with any potential model employed. We suggest to use a
phase equivalent transformation of the two-body interaction  to
improve the description of the  $^6$He binding. If the phase
equivalent transformation depends on a continuous parameter(s) than
varying the parameter we can  fit the
$^6$He binding energy to the phenomenological value. We suppose this
approach to be interesting for various few-body applications.

Various phase equivalent transformations have been discussed in
literature. The local phase equivalent transformations (transforming a
local potential into another local potential  phase equivalent to the
initial one) is well-known (see, e.~g., Ref.~\cite{Newton}). In the
inverse scattering theory, this transformation gives rise to 
the ambiguity of the potential restored from the scattering
data. Recently this transformation was extended on the case of the
multichannel scattering~\cite{Sid}. However the local phase equivalent
transformation can be applied only to a two-body system that have at
least one bound states and 
the number of continuous parameters of the transformation is equal to
the number of the bound states in the system. Therefore this
transformation cannot be applied to the $n+\alpha$ system that has no
bound states.

Recently the so-called supersymmetry phase equivalent transformation
(see, e.~g., review \cite{SUSY-rev}) become very popular. Using this
transformation one can transform a potential with the Pauli forbidden
state into exactly  phase equivalent potential with additional
repulsion simulating the Pauli effects. The effect of this
transformation on the properties of three-body cluster systems was
examined in a number of recent papers (see, e.~g.,
Ref.~\cite{Fed-SUSY}). The  supersymmetry  transformation was shown
\cite{Sid} to be a particular case of the local phase equivalent
transformation. However the  supersymmetry  transformation does not
have parameters and cannot be used for our purposes.

A phase equivalent transformation based on unitary transformation
of the Hamiltonian, was suggested in Ref.~\cite{matter}. This
transformation has continuous parameters. This transformation have
been used in many-body applications in
Ref.~\cite{Vary-PHT}. However the authors of Ref.~\cite{Vary-PHT} used
few particular cases of the transformation corresponding to particular
parameter values and did not try to vary the parameters continuously.

We have developed recently \cite{PHT} a phase equivalent
transformation using the $J$-matrix formalism. Generally, our
transformation is a particular case of the  phase equivalent
transformation of Ref.~\cite{matter}. However we suppose that our
phase equivalent transformation is general enough and very convenient
for the use in various many-body applications utilizing any $L^2$
basis. We applied this transformation to the $NN$ interaction and
studied the effect of the transformation on the properties $^3$H and
$^4$He nuclei in Ref.~\cite{SVWM}.
In what follows, we discuss briefly the transformation and its
application to the $^6$He nucleus within the three-body cluster model.

We suppose that a two-body system is described by the Schr\"odinger
equation
\begin{gather}
H\Psi(r)=E\Psi(r),
\label{Schr-2}
\end{gather}
where the Hamiltonian $H = T + V$,
$T$  and $V$ are kinetic and  potential energy operators.
Introducing a complete basis $\{\phi_{\kappa}(r)\}$, 
$\kappa=0$, 1, 2,~... of $L^2$
functions $\phi_\kappa(r)$, we can expand the 
solutions of Eq.~(\ref{Schr-2})  in infinite series of basis functions
$|\kappa\rangle\equiv\phi_\kappa(r)$: 
\begin{equation}
\Psi(r) = \sum\limits_{\kappa=0}^{\infty} C_{\kappa} |\kappa\rangle  .
\label{PhEq_1}
\end{equation}
The Schr\"odinger equation (\ref{Schr-2}) takes the form of an infinite
dimensional algebraic problem
\begin{equation}
\sum_{\kappa'=0}^\infty \langle\kappa|H|\kappa'\rangle\,C_{\kappa'}=EC_\kappa,
\label{Al-inf}
\end{equation}
where $\langle\kappa|H|\kappa'\rangle$ are the matrix elements of the  infinite
dimensional Hamiltonian matrix $\left[H\right]$.

Now we define a new matrix
\begin{equation}
\left[H'\right] = \left[U^{+}\right]\,\left[H\right]\,\left[U\right] 
  \label{PhEq1}
\end{equation}
with the help of the unitary matrix $\left[U\right]$ which is supposed
to be of the  form
\begin{equation}
\left[U\right] = \left[U_{0}\right]  \oplus  \left[I\right]
            = \left(\begin{array}{c|c}
                          \left[U_{0}\right] &  0  \\ \hline
                               0             & \left[I\right]
                       \end{array}\right)  , 
\label{PhEq2}
\end{equation}
where $\left[I\right]$ is the infinite dimensional unit matrix and
$\left[U_{0}\right]$ is $N\times N$ unitary matrix. A new Hamiltonian $H'$ is
defined through its matrix $\left[H'\right]$. It is supposed that
$\left[H'\right]$ is the matrix of the 
Hamiltonian $H'$ in the {\em original} basis  $\{\phi_\kappa(r)\}$. 

Clearly the spectra of Hamiltonians $H$ and $H'$ are identical. The
difference between the
eigenfunctions $\Psi'(r)$ of the Hamiltonian ${H}'$ and the
eigenfunctions $\Psi(r)$ of the Hamiltonian ${H}$ corresponding to the
same energy $E$,  is a
superposition  of a finite number of square integrable functions: 
\begin{equation}
\Psi'(r) = \Psi(r) +
\sum\limits_{\kappa=0}^{N-1} \Delta C_{\kappa} \: \phi_{\kappa}(r) .
\label{wf-PET}
\end{equation}
The superposition of  a finite number of $L^2$ functions cannot affect
the asymptotics of scattering wave functions. Since the scattering
phase shifts and the $S$-matrix are defined through the asymptotic
behavior of the wave functions, the phase shifts associated with the
functions  $\Psi(r)$ and  $\Psi'(r)$ are identical. In other words,
the Hamiltonians  ${H}$  and $H'$ are phase equivalent.

Supposing that the Hamiltonian $H'=T+V_{\rm PET}$, we introduce  the potential
\begin{gather}
V_{\rm PET}=V+\Delta V
\label{PEQ-V}
\end{gather}
phase equivalent to the original potential $V$. The potential $V_{\rm PET}$ is
defined through its matrix
\begin{gather}
\left[V_{\rm PET}\right]=\left[V\right]+\left[\Delta V\right]
\label{matrPEQ-V}
\intertext{in the  basis  $\{\phi_\kappa(r)\}$, where}
\left[\Delta {V}\right] 
=\left[ {U}^{+}\right] \,\left[{H}\right]\,\left[ {U}\right]-\left[{H}\right] .
\label{matrPEQ-DV}
\end{gather}

The net result of the above considerations is very simple. We
introduce any complete $L^2$ basis  $\{\phi_n(r)\}$
and calculate the Hamiltonian
matrix in this basis. Next we introduce an arbitrary unitary
transformation of the type (\ref{PhEq2}) that affects a finite number
of the basis functions. With the help of this transformation, we
calculate the additional potential $\Delta V$ using
Eq.~(\ref{matrPEQ-DV}) and obtain the phase equivalent interaction by
means of Eq.~(\ref{PEQ-V}). The additional potential $\Delta V$ is
non-local, and hence the  phase equivalent potential $V_{\rm PET}$ is
non-local, too. Therefore the suggested transformation is a
{\em non-local  phase equivalent transformation}: 
it brings us to a non-local interaction
phase equivalent to the original one. 

The suggested non-local  phase equivalent transformation can be easily
implemented in many-body calculations utilizing any $L^2$ basis. One
just needs to add the two-body kinetic energy matrix to the two-body
interaction matrix, unitary transform the obtained matrix and subtract
the  kinetic energy matrix from the result to obtain the matrix of the
phase equivalent two-body interaction. 

The  non-local  phase equivalent transformation can be easily
explained and understood within the $J$-matrix formalism. In the
$J$-matrix formalism, the $S$-matrix  and the phase shifts are governed by
the matrix elements\linebreak 
${\langle \kappa_{\Gamma\vphantom{'}}\Gamma 
|\mrs P|\kappa_{\Gamma'\vphantom{'}}+1,\Gamma'\rangle}$
 which are defined through eigenvalues
$E_\lambda$ and the last component
$\langle\kappa_{\Gamma\vphantom{'}}\Gamma|\lambda \rangle$  of the
eigenvectors  of the
truncated Hamiltonian matrix [see
Eqs.~(\ref{e43})-(\ref{e44})]. If the truncated Hamiltonian matrix is
larger than the non-trivial submatrix  
$\left[U_{0}\right]$ of the unitary matrix $\left[U\right]$, than
neither $E_\lambda$ nor the  last component
$\langle\kappa_{\Gamma\vphantom{'}}\Gamma|\lambda \rangle$  of the
eigenvectors are affected by the unitary transformation
(\ref{PhEq2}). Therefore neither  the $S$-matrix nor the phase shifts
are affected by the transformation. Nevertheless the wave function is
seen from Eq.~(\ref{wf-PET}) to be subjected to changes by the
transformation; in other words, the off-shell properties of the
original and the transformed potentials are different. The non-local
phase equivalent transformation generates the ambiguity of the
interaction obtained by means of the $J$-matrix version of the inverse
scattering theory \cite{Z-inv}.

The off-shell properties of the two-body interactions play an
important role in the formation of the  properties of a three-body
(and generally many-body) system. Therefore it is interesting to
investigate the effect of the non-local phase equivalent
transformation on the properties of  the
$^6$He nucleus within the  three-body cluster model. We note that the
non-local phase equivalent transformation preserves the energies of
the two-body bound states. However the properties of these states are
not preserved, for example, the rms radius of the  two-body bound
system may be affected by the transformation~\cite{Polyz}. Hence it is somewhat
dangerous to apply the transformation to the $NN$ interaction since it
may destroy the description of the deuteron properties. We have
already noted that we employ effective $n{-}n$ interactions that are 
designed for the effective description of many-body nuclear systems
and that are not supposed to be careful in the description of the
dinucleon. Nevertheless we suppose that it is more natural to apply
the transformation to the $n{-}\alpha$ interaction since the
$n+\alpha$ system does not have a bound state and all the information
about this interaction is extracted from the scattering data only. It is
clear from the Table~\ref{He6_weights}\,(c) that the $^6$He properties
are dominated by the $p$ wave component of  the $n{-}\alpha$
interaction. Therefore we apply the non-local phase equivalent
transformation to  the $p$ wave component of  the $n{-}\alpha$
potential only.

The non-local phase equivalent transformation for  the $n{-}\alpha$
potential  $p$ wave component is constructed in the oscillator basis
with $\hbar\omega$ values used in three-body calculations with the respective
potential model (see Table~\ref{He6_energies_radii}). The simplest
non-local transformation involves a $2\times 2$ unitary matrix
$\left[U_{0}\right]$. Any $2\times 2$ unitary matrix is known to be a
rotation matrix with a single continuous parameter~$\gamma$:
\begin{equation}
\left[U_{0}\right]=\begin{bmatrix}
  \cos\gamma & -\sin\gamma \\
             +\sin\gamma & \cos\gamma    
\end{bmatrix} .  
\label{PhEqTr2}
\end{equation}
A more complicated non-local transformation involves a $3\times 3$
rotation matrix $\left[U_{0}\right]$ with two continuous parameters
$\gamma$ and $\beta$: 
\begin{equation}
\left[U_{0}\right]=\begin{bmatrix}
\cos\beta\, \cos\gamma
                & -\cos\beta\, \sin\gamma \ \ &  \sin\beta  \\
\sin\gamma &  \cos\gamma     &  0    \\
 -\sin\beta\, \cos\gamma &  \sin\beta\, \sin\gamma  & \cos\beta
\end{bmatrix} .          
\label{PET_2D}
\end{equation}
Clearly the transformation with the matrix (\ref{PET_2D}) is
equivalent to the transformation with the matrix (\ref{PhEqTr2}) if  
the Euler angle $\beta=0$.

\begin{figure}
\centerline{\hspace{-5pt}%
\includegraphics[width=0.5\textwidth,angle=0]{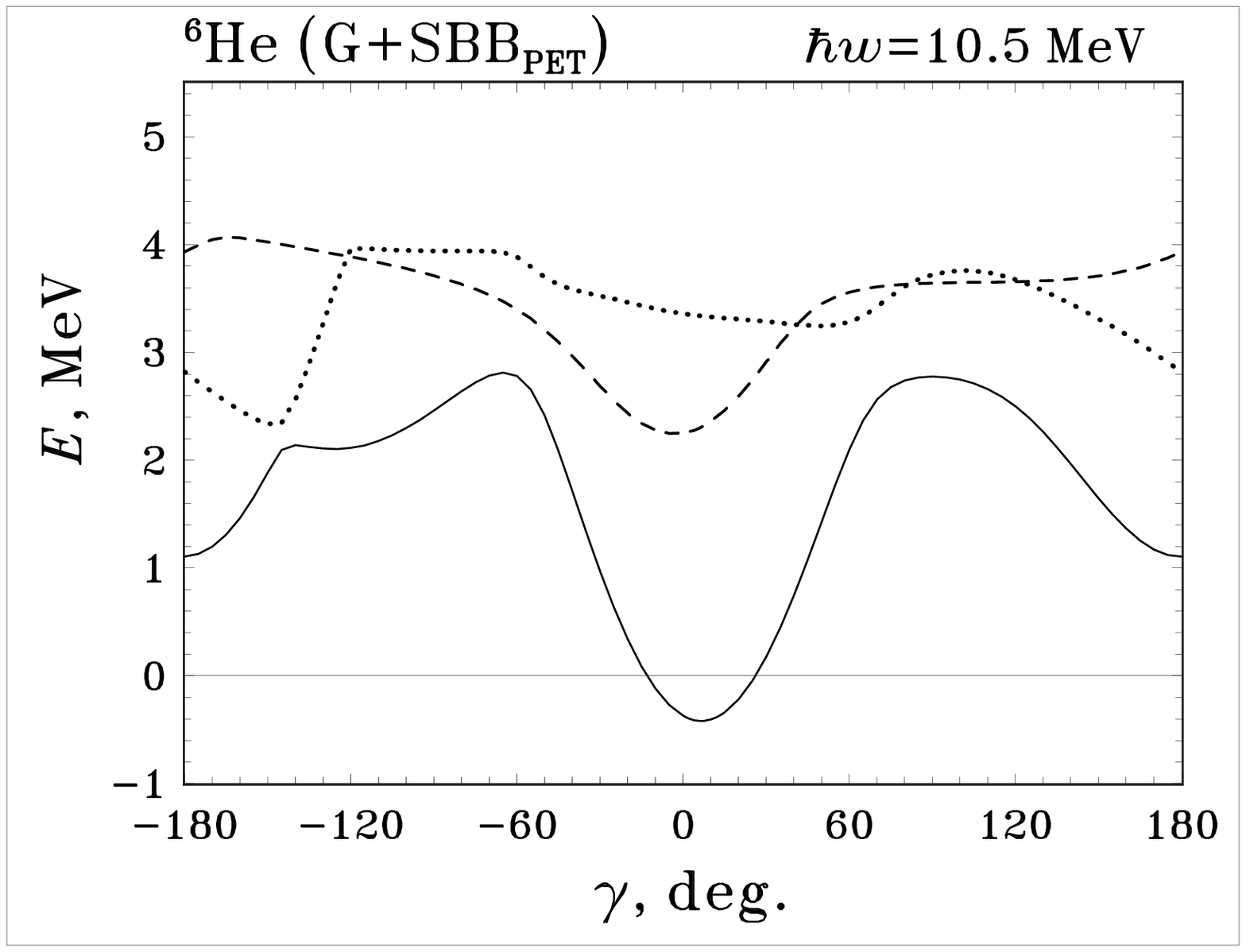}
\hfill
\includegraphics[width=0.5\textwidth,angle=0]{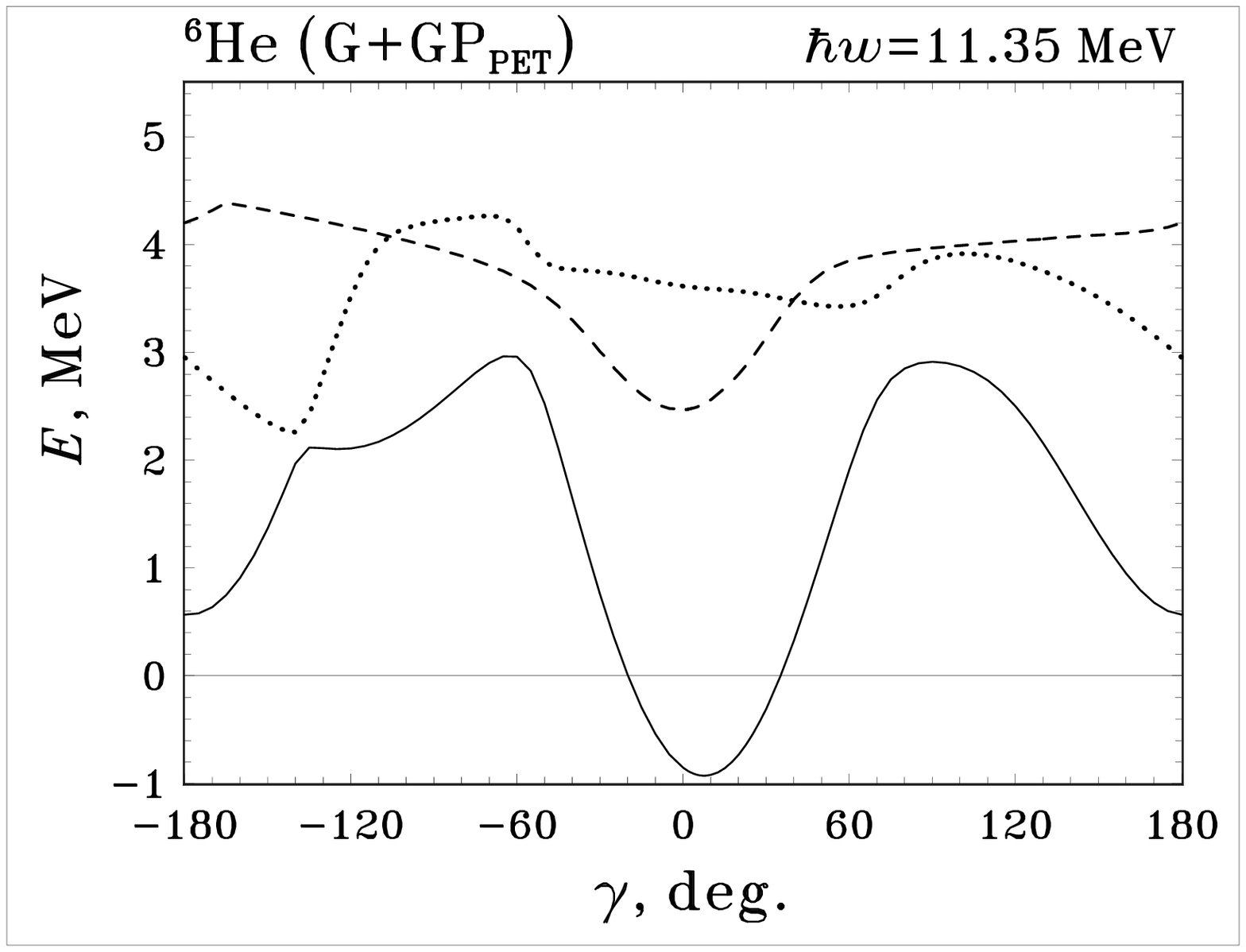}
                   }\vspace{10pt}
\centerline{\hspace{-5pt}%
\includegraphics[width=0.5\textwidth,angle=0]{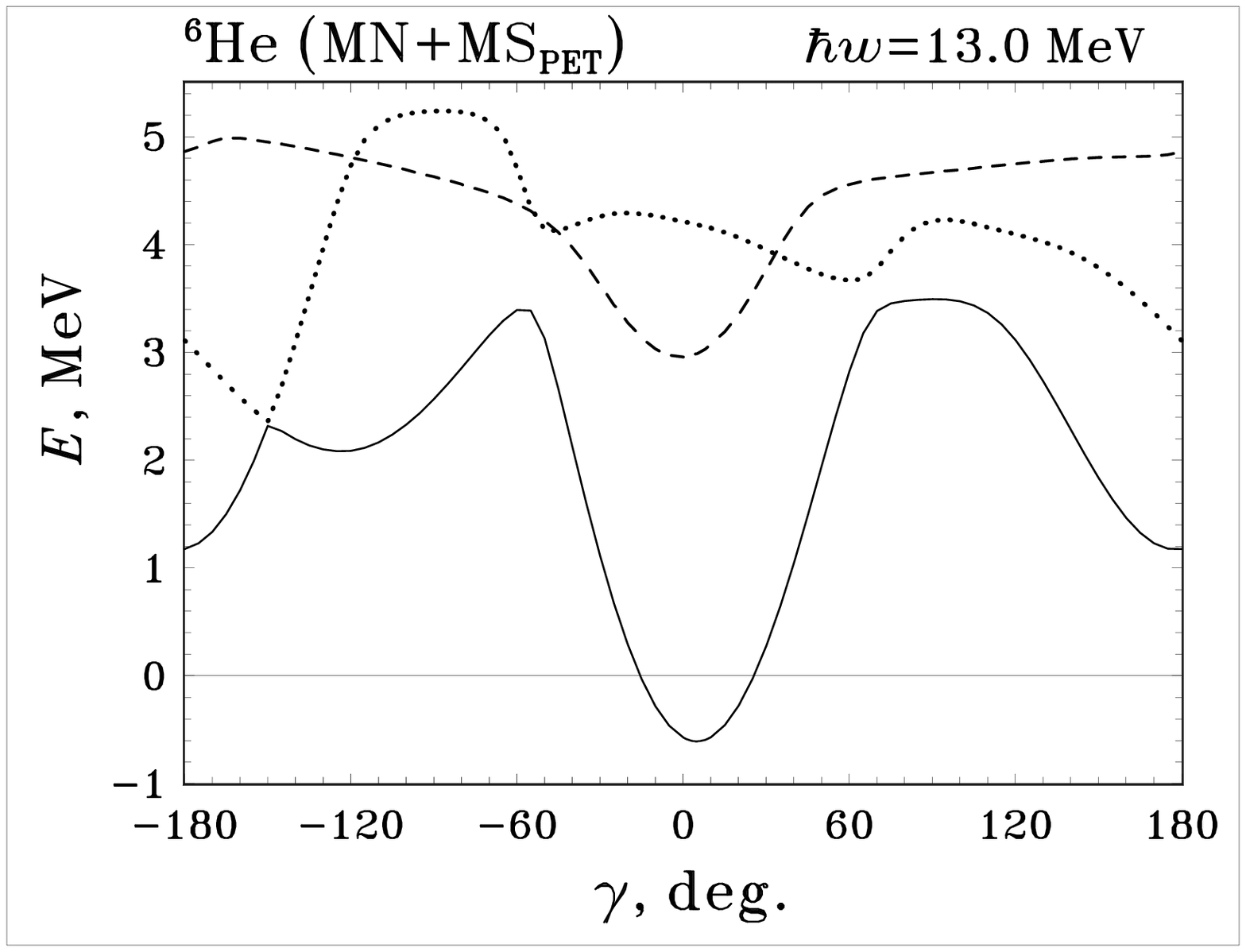}
\hfill
\includegraphics[width=0.5\textwidth,angle=0]{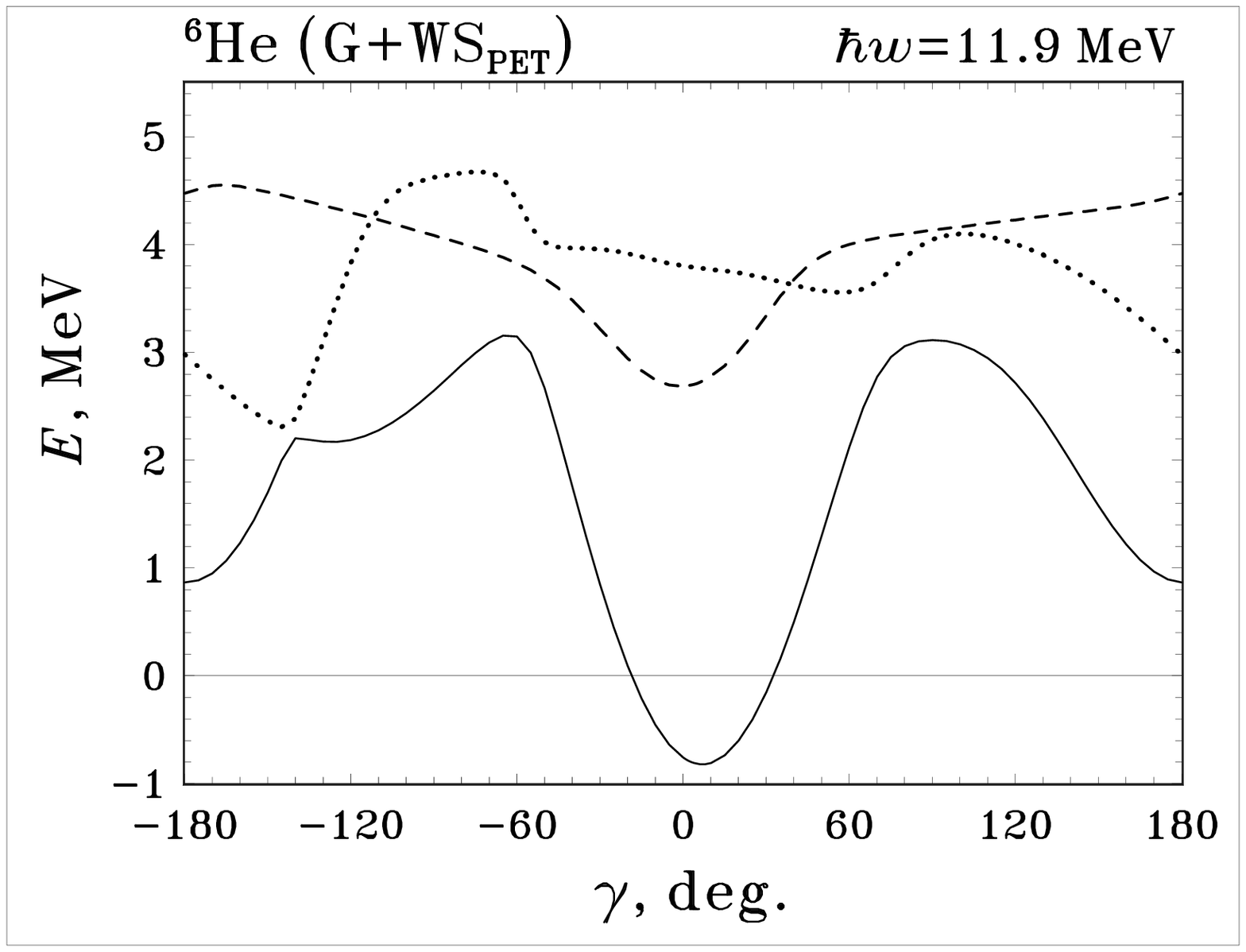}
                   }

\caption{
The $^6$He ground state energy (solid line) and the energies of the
lowest  $1^-$ (dashed line)  and $0^+$ (dotted line)  excited states
obtained in the variational approach with phase equivalent
$n{-}\alpha$ potentials vs the transformation parameter $\gamma$ of the
one-parameter  transformation
(\ref{PhEqTr2}). 
The truncation parameter $\tilde{N}=20$ in the $0^+$ state calculations
and $\tilde{N}=21$ in the $1^-$ state calculations.
}
\label{fig_PET_energy}
\end{figure}

We present in Fig.~\ref{fig_PET_energy} the results of variational
calculations of the  $^6$He ground and the first excited $0^+$ states with
$\tilde{N}=20$  and of the lowest $1^-$ state with  $\tilde{N}=21$
with various $n{-}n$ potentials and $n{-}\alpha$ potentials
obtained by applying the phase equivalent transformation
(\ref{PhEqTr2}) to  various original potentials. The $\gamma$
dependence of $0^+$ and $1^-$ state energies is very interesting. It
is seen from  Fig.~\ref{fig_PET_energy} that the simplest non-local
phase equivalent transformation of $p$ wave component of the
$n{-}\alpha$ potential  only, can completely change the spectrum of
the three-body cluster system. The $\gamma$ dependence of the energies
is seen to be very similar for all potential models under consideration. 
The  variations of the energies of the excited $0^+$
and $1^-$ states are seen to be smaller than the ground state energy
variations. There is the $0^+$ states level crossing at
$\gamma\approx -150^\circ$ for all potential model considered.

\begin{figure}
\centerline{\hspace{-5pt}%
\includegraphics[width=0.5\textwidth,angle=0]{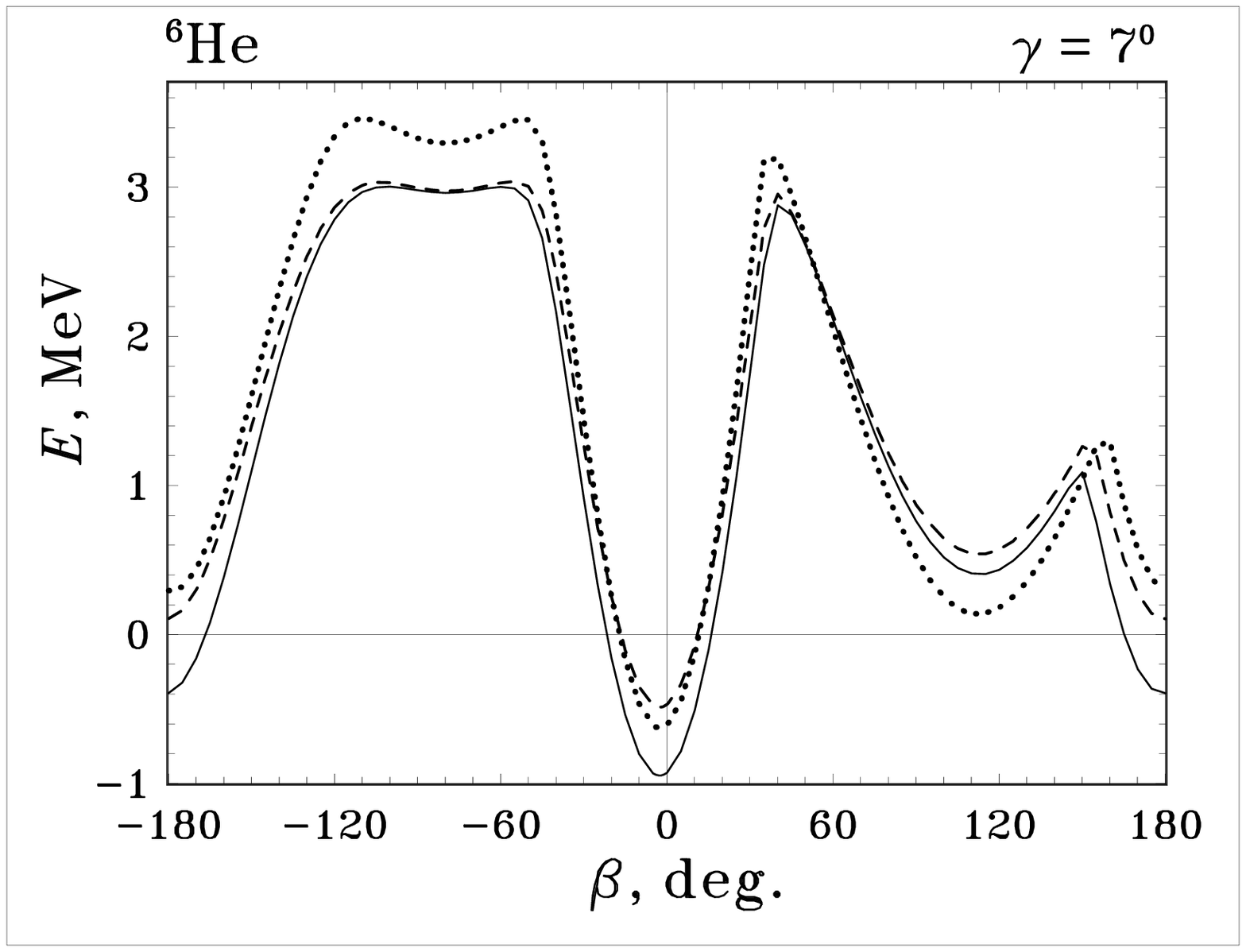}
\hfill
\includegraphics[width=0.5\textwidth,angle=0]{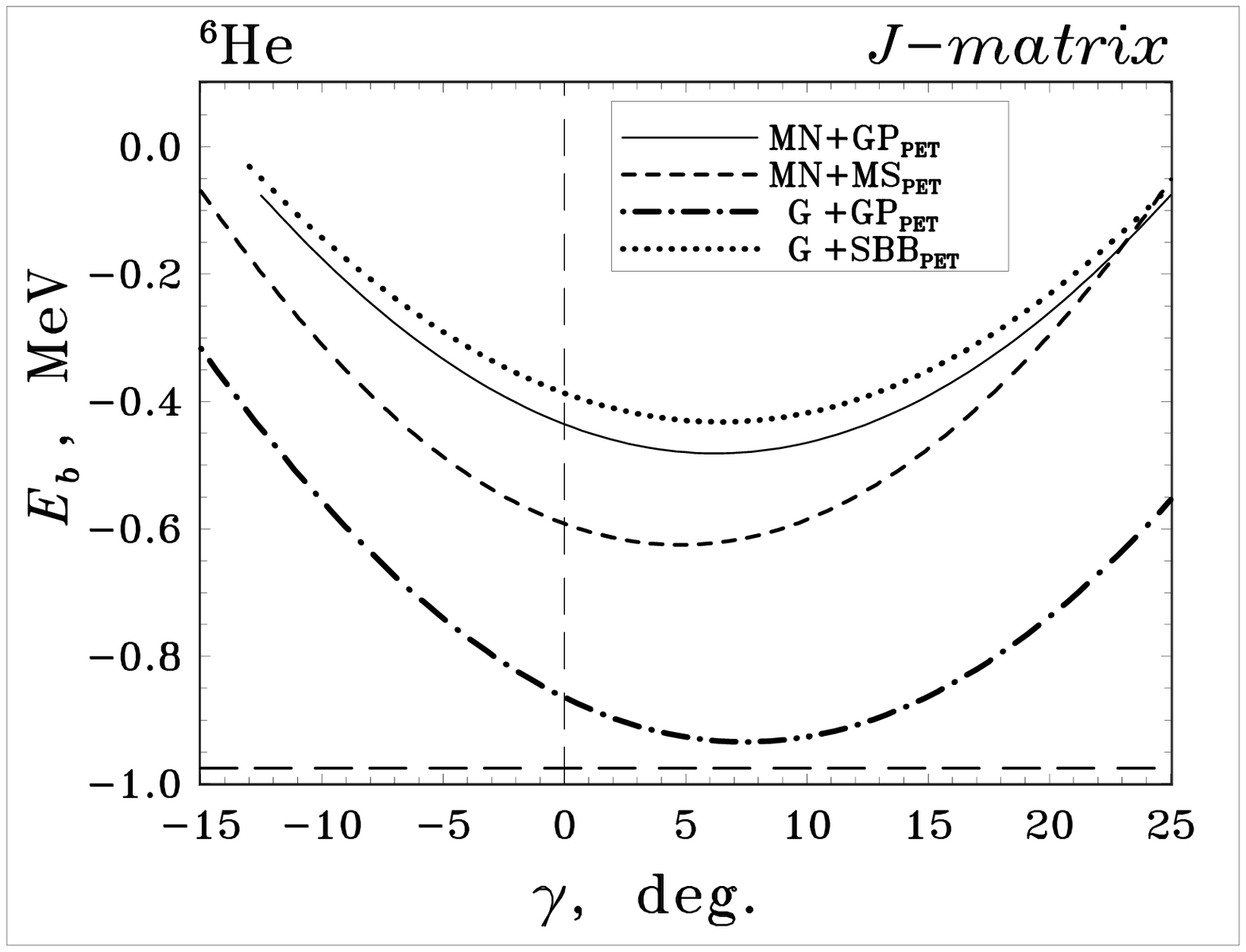}
}

\parbox[t]{0.48\textwidth}{
\caption{
The $^6$He ground state energy 
vs the transformation parameter $\beta$ of the phase equivalent transformation
(\ref{PET_2D}) with $\gamma=7^\circ$.
Solid, dashed and dotted lines are calculations in the variational
approximation with $\rm G+GP_{\rm PET}$, 
$\rm MN+GP_{\rm PET}$ 
and $\rm MN+MS_{\rm PET}$ potential models, respectively, with
$\hbar\omega$ values given in Table~\ref{He6_energies_radii} and   
$\tilde N=20$.
        }
\label{fig_PET_2D}
}
\hfill
\parbox[t]{0.48\textwidth}{
\caption{
The $^6$He ground state energy 
obtained in the $J$-matrix approach with phase equivalent
$n{-}\alpha$ potentials vs the transformation parameter $\gamma$ of the
one-parameter  transformation (\ref{PhEqTr2}). 
The truncation parameter $\tilde{N}=20$, the $\hbar\omega$ values for
each potential model can be found in 
Table~\ref{He6_energies_radii}. The straight dashed line is the
experimental ground state energy.
        }
\label{fig_PET_jmatr}
}

\end{figure}

The ground state energy $\gamma$ dependence passes through a minimum in the
vicinity of $\gamma=5{-}7^{\circ}$. The minimum corresponds to the increase of
the binding by approximately $0.04{-}0.07$~MeV or $7{-}12$\% depending
on the potential model, 
that improves the results obtained with
the original potentials ($\gamma=0$) given in
Table~\ref{He6_energies_radii}. The ground states may be additionally
shifted down using  the two-parameter transformation
(\ref{PET_2D}). Supposing $\gamma=7^\circ$ and varying the parameter
$\beta$ we obtain variational results shown in
Fig.~\ref{fig_PET_2D}. The ground state $\beta$ dependence has a
minimum at $\beta\approx -2.5^\circ$ for all potential models. This
minimum corresponds to the additional binding of
$0.019\div 0.023$~MeV (about 2--4\%).

The variational results presented in 
Figs.~\ref{fig_PET_energy} and  \ref{fig_PET_2D}  are interesting for
understanding the general trends of variation of the ground state
energy and $^6$He spectrum when the parameters of the phase equivalent
transformations (\ref{PhEqTr2})--(\ref{PET_2D}) are varied in a wide
range of values. The corrections to the $^6$He binding due to the
effect of the phase equivalent transformation  are
better seen in Fig.~\ref{fig_PET_jmatr} where we present in a larger
scale the $J$-matrix results for the ground state energies. The $^6$He
binding energy is seen to be very close to the empirical value in the
$\rm G+GP_{PET}$ potential model when $\gamma\approx 7^\circ$. 
However
the results presented in  Fig.~\ref{fig_PET_jmatr} were obtained with
not very large value of the truncation parameter $\tilde N=20$. With
$\tilde N=28$ and $\gamma=7.5^\circ$ we obtain in this potential model
an excellent description of the $^6$He binding energy $E_b=0.952$~MeV and 
rms radius $\langle r^2\rangle^{1/2}=2.37$~fm. Using the two-parameter
transformation (\ref{PET_2D}) with $\gamma=7.5^\circ$ and
$\beta=-3^\circ$, we obtain $E_b=0.973$~MeV and 
$\langle r^2\rangle^{1/2}=2.36$~fm, i.~e. the phenomenological value
$E_b=0.976$~MeV is reproduce nearly exactly. 

\begin{figure}
\centerline{
\includegraphics[width=0.6\textwidth,angle=0]{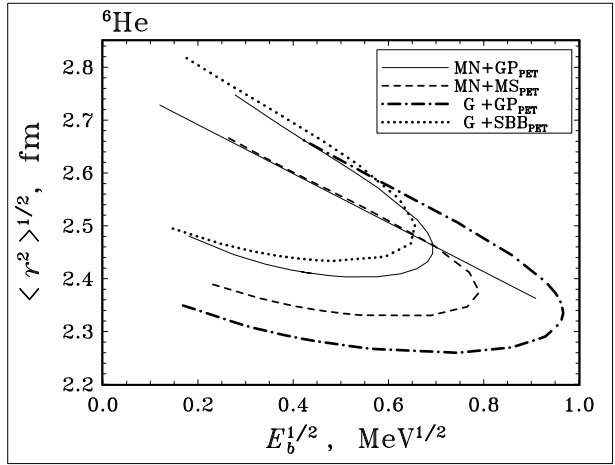}
}

\caption{
Correlations between $E_b^{1/2}$ and rms radius in $^6$He obtained by
variation of the parameter $\gamma$ of the phase equivalent
transformation  (\ref{PhEqTr2}) in various potential models
($J$-matrix approximation with  $\tilde{N}=20$, the corresponding
$\hbar\omega$ values   can be found in 
Table~\ref{He6_energies_radii}).
A straight solid line is added to visualize the linear
correlation between
$\langle r^2\rangle^{1/2}$ and $E_b^{1/2}$ on a part of the trajectory.
        }
\label{fig_PET_rad_enr}

\end{figure}

The correlations between the $^6$He rms radius $\langle r^2\rangle^{1/2}$
and the square root of the
binding energy~$E_b^{1/2}$, are depicted in
Fig.~\ref{fig_PET_rad_enr}. We present the $J$-matrix results obtained
with various potential models when the parameter $\gamma$ of the phase
equivalent transformation (\ref{PhEqTr2}) varies on the interval where
$^6$He appears to be bound in the variational approximation in
the given potential model.
The correlations are seen 
to be very interesting: there are two very different 
$\langle r^2\rangle^{1/2}$ values that are in  correspondence with 
the same binding energy. For $\gamma\lesssim -6^\circ$, the rms radius
 $\langle r^2\rangle^{1/2}$ decreases linearly with $E_b^{1/2}$.

\begin{figure} [b]
\centerline{\hspace{-5pt}%
\includegraphics[width=0.5\textwidth,angle=0]{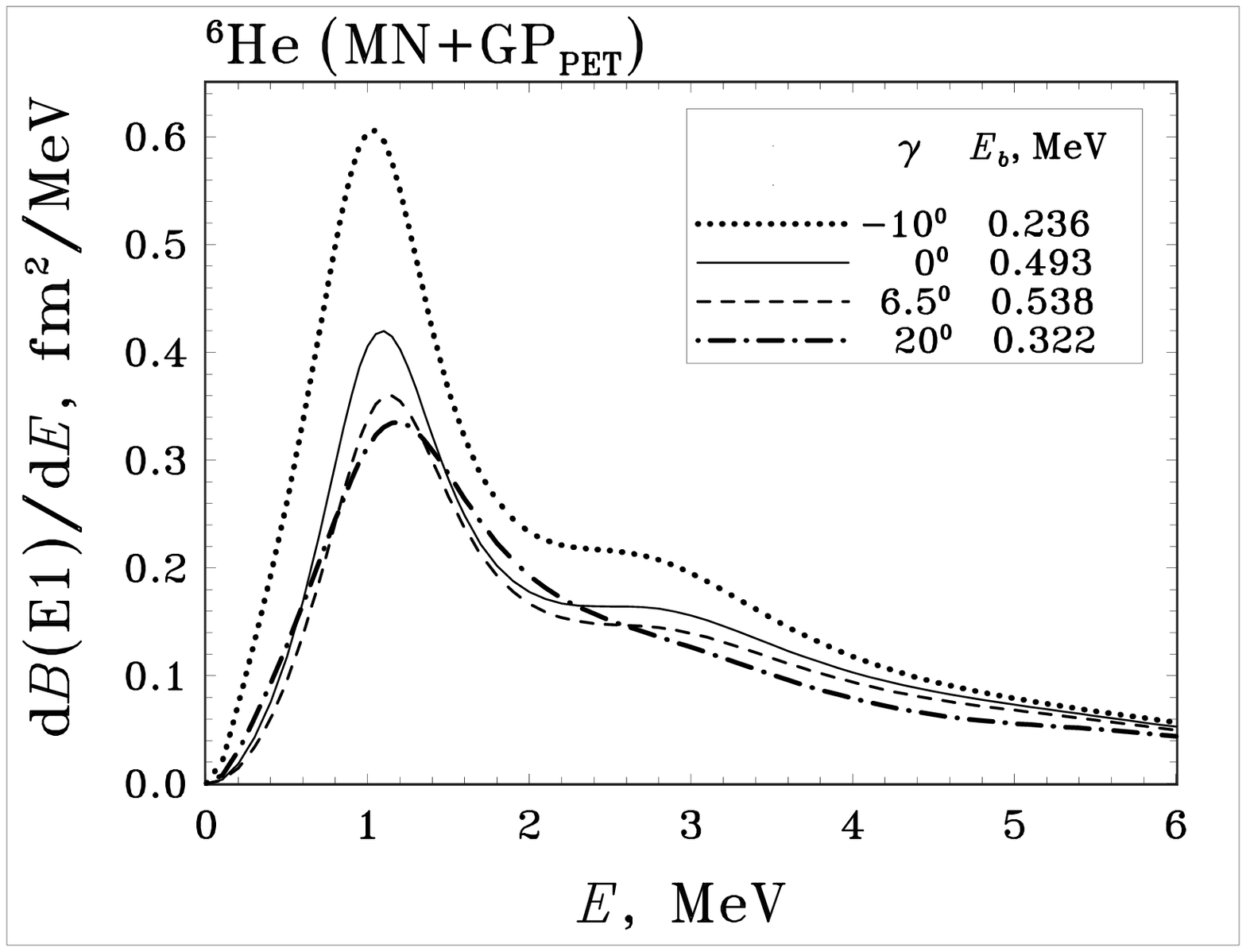}
\hfill
\includegraphics[width=0.5\textwidth,angle=0]{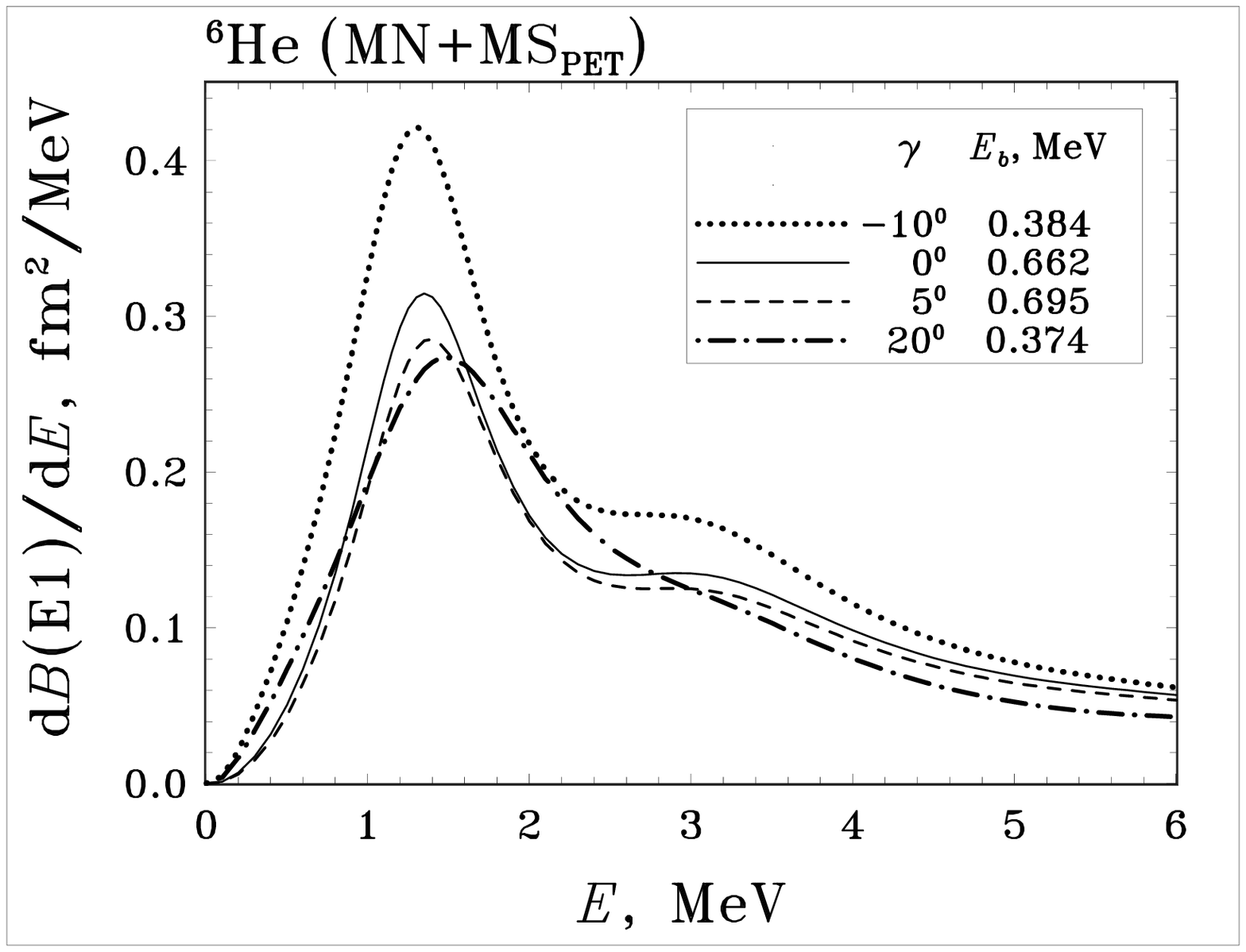}
}

\caption{
Reduced $E1$ transition probability  
$\frac{d\,{\mathcal B}(E1)}{d\,E}$ in  $^{6}$He obtained 
in the JJ approximation with the ground state 
truncation parameter $\tilde{N}_{\rm g.s.}=22$ and the final   state 
truncation parameter $\tilde{N}_{\rm f.s.}=23$ with the 
${\rm MN+GP_{PET}}$ (left panel) and ${\rm MN+MS_{PET}}$ (right panel)
potential models (the corresponding  
$\hbar \omega$ values can be found in  Table~\ref{He6_energies_radii})
and different values of the  parameter $\gamma$ of the phase equivalent
transformation~(\ref{PhEqTr2}) (the corresponding binding energies
are listed in the figure).
        }
\label{fig_PET_B(E)}

\end{figure}

The effect of the phase equivalent transformation  (\ref{PhEqTr2}) on the 
reduced $E1$ transition 
probability $\displaystyle\frac{d\,{\mathcal B}(E1)}{d\,E}\vphantom{\int}$ 
in  $^{6}$He, is
illustrated by  Fig.~\ref{fig_PET_B(E)}. The effect is seen to be
essential. The naive expectation is that the soft dipole mode will
be more enhanced if the binding energy is smaller. However the
strength of the $E1$ transitions in the vicinity of the maximum of   
$\displaystyle\frac{d\,{\mathcal B}(E1)}{d\,E}\vphantom{\int^A}$ does not
demonstrate so 
simple dependence on the binding energy $E_b$ (the corresponding $E_b$
values are listed in the figure).


\section{Conclusions}

The hyperspherical $J$-matrix formalism  makes it possible to
study not only the few-body disintegration  of the system but also to
calculate the $S$-matrix poles in the $A$-body system. This approach
appears to be a very powerful tool for calculations of the bound
state properties. As a result, we obtain a unified theory that is
capable to investigate in a unique approach both the discrete and
continuum spectra of $A$-body systems.

The suggested $J$-matrix motivated non-local phase equivalent
transformation may be used to fit binding energies of many-body
systems in the case when all information about two-body interaction is
extracted from the scattering data only. The transformation can be
easily utilized  
in the studies of many-body systems 
with any $L^2$ basis.

\bigskip 

This work was supported in part  by the State Program ``Russian
Universities'' and by the Russian Foundation of Basic Research, Grant 
No~02-02-17316.

\end{document}